\documentclass[12pt]{spieman}  
\usepackage{amsmath,amsfonts,amssymb}
\usepackage{graphicx}
\usepackage{setspace}
\usepackage{tocloft}

\usepackage{multirow}
\usepackage{lineno}
\title{In-flight calibration system of Imaging X-ray Polarimetry Explorer}

\author[a,b*]{Riccardo Ferrazzoli}
\author[a]{Fabio Muleri}
\author[a]{Carlo Lefevre}
\author[a]{Alfredo Morbidini}
\author[a]{Fabrizio Amici}
\author[a]{Daniele Brienza}
\author[a]{Enrico Costa}
\author[a]{Ettore Del Monte}
\author[a]{Alessandro Di Marco}
\author[a]{Giuseppe Di Persio}
\author[c]{Immacolata Donnarumma}
\author[a]{Sergio Fabiani}
\author[a]{Fabio La Monaca}
\author[a]{Pasqualino Loffredo}
\author[d]{Luca Maiolo}
\author[d]{Francesco Maita}
\author[a]{Raffaele Piazzolla}
\author[e]{Brian Ramsey}
\author[a,f]{John Rankin}
\author[a,f]{Ajay Ratheesh}
\author[a]{Alda Rubini}
\author[g]{Paolo Sarra}
\author[a]{Paolo Soffitta}
\author[a]{Antonino Tobia}
\author[a]{Fei Xie}

\affil[a]{INAF-IAPS, via del Fosso del Cavaliere 100, 00133 Roma, Italy}
\affil[b]{Universit\`a di Roma "Sapienza",Dipartimento di Fisica, Piazzale Aldo Moro 5, 00185 Roma, Italy}
\affil[c]{ASI, Via del Politecnico snc, 00133 Roma, Italy}
\affil[d]{CNR - IMM, via del Fosso del Cavaliere 100, 00133 Roma, Italy}
\affil[e]{NASA Marshall Space Flight Ctr., Huntsville, AL 35812, USA}
\affil[f]{Universit\`a di Roma Tor Vergata, Dipartimento di Fisica, via Cracovia 50 1,00133 Roma, Italy}
\affil[g]{OHB Italia SpA, Via Gallarate 150, 20151 Milano, Italy}

\cftpagenumbersoff{figure}
\cftpagenumbersoff{table}

\begin{document} 
\maketitle

\begin{abstract}
The NASA/ASI Imaging X-ray Polarimetry Explorer, which will be launched in 2021, will be the first instrument to perform spatially resolved X-ray polarimetry on several astronomical sources in the 2--8 keV energy band.
These measurements are made possible owing to the use of a gas pixel detector (GPD) at the focus of three X-ray telescopes.
The GPD allows simultaneous measurements of the interaction point, energy, arrival time, and polarization angle of detected X-ray photons. 
The increase in sensitivity, achieved 40 years ago, for imaging and spectroscopy with the Einstein satellite will thus be extended to X-ray polarimetry for the first time.\\
The characteristics of gas multiplication detectors are subject to changes over time. 
Because the GPD is a novel instrument, it is particularly important to verify its performance and stability during its mission lifetime. \\
For this purpose, the spacecraft hosts a filter and calibration set (FCS), which includes both polarized and unpolarized calibration sources for performing in-flight calibration of the instruments.
In this study, we present the design of the flight models of the FCS and the first measurements obtained using silicon drift detectors and CCD cameras, as well as those obtained in thermal vacuum with the flight units of the GPD.\\
We show that the calibration sources successfully assess and verify the functionality of the GPD and validate its scientific results in orbit; this improves our knowledge of the behavior of these detectors in X-ray polarimetry.
\end{abstract}


{\noindent \footnotesize\textbf{*}Riccardo Ferrazzoli,  \linkable{riccardo.ferrazzoli@inaf.it} }


\section{Introduction}
\label{sect:intro}  
The Imaging X-ray Polarimetry Explorer (IXPE \cite{2016SPIEWeisskopf}) will be the first mission entirely dedicated to X-ray polarimetry, scheduled to be launched in 2021. 
In addition to the direction, energy, and arrival time of every photon, polarimetry adds two observables: the degree and angle of polarization.
These observables provide information on the emission mechanism and the geometry of the source. 
To date, only few experiments conducted X-ray polarimetry of celestial sources \cite{1972Novick,1976Weisskopf,1978Weisskopf_POL}.
They were extensively limited in their observation time and by the competition with other instruments onboard the space observatories, which resulted in the field staying dormant for decades.
IXPE was first proposed in response to an Announcement of Opportunity issued in 2014, where it was selected in the context of the NASA Astrophysics Small Explorer (SMEX) program in January 2017 in collaboration with the Italian Space Agency (ASI). 
IXPE comprises three detector units (DUs), each containing a gas pixel detector X-ray imaging polarimeter (GPD).
The innovative technology of the GPD significantly increases the polarimetric sensitivity compared to that of the X-ray polarimeters flown to date, simultaneously measuring the X-ray radiation interaction point, energy, arrival time, and polarization angle.
This enables precise X-ray polarimetric measurements of numerous classes of cosmic sources, ranging from neutron stars to black hole binaries and active galactic nuclei, as well as extended sources such as supernova remnants, pulsar wind nebulae, and large-scale jets.
After the first development phase \cite{2001Costa_POL,2004Bellazzini,2006Bellazzini,2007Bellazzini}, the prototype GPD has been extensively tested for more than 15 years. 
However, its complexity requires accurate monitoring of its performance during the mission lifetime. 
Generally, the characteristics of detectors based on gas multiplication change on the time scale of several years, thus requiring a detailed plan for in-orbit calibrations.
The main motivation behind this mission is to deliver ground-breaking scientific observations; however, operating and calibrating the GPD in orbit will pose a serious challenge.
As of today, only one source, the Crab Nebula \cite{1978Weisskopf_POL}, is known to emit polarized X-rays. 
However, it cannot be used as a polarized calibration source, as its flux is changing with time and the effect of this on its polarization is unknown.
Moreover, one of the scientific motivations of IXPE is the possibility of detecting polarization degrees down to a $\sim$ 1\% level, and because virtually all celestial sources (with exception maybe of clusters of galaxies) are expected to be polarized at some level which is not known a priori, in order to calibrate in orbit the detector no astrophysical object can be used as unpolarized calibration source.
This ultimately leads to the stringent requirement of calibrating with sources installed in the on-board instrumentation.
To this end, during the mission lifetime, the GPD response will be monitored using a filter and calibration set (FCS) hosted on a filter and calibration wheel (FCW) included in each DU \cite{2018Muleri}.
Each FCS consists of four calibration sources, namely, CalA, CalB, CalC, CalD, and filters for special observations.
The sets are denoted as flight model (FM)1, FM2, FM3, and FM4, and they are assigned to the three DUs that will fly onboard the IXPE spacecraft and one DU that will act as a spare.
The FCS includes both polarized and unpolarized calibration sources, capable of illuminating the whole detector or just a part of it, for mapping and monitoring of the GPD modulation factor (i.e., the detector response in terms of modulation to 100\% polarized radiation), quantum efficiency, and energy resolution at different energies.
In-orbit calibrations will also allow us to check for the presence of spurious polarization, as well as to map and monitor the gain and its uniformity across the 15$\times$15 mm$^2$ detector surface. \\
These pieces of information will help improve our understanding of the detector performance and asses the reliability of scientific results. \\  
In this study, we report the results of the first set of measurements performed on the four FMs of the polarized and unpolarized in-flight calibration sources. 
The paper is organized as follows. 
In Sec.~(\ref{sec:methods}), we describe the calibration sources and the experimental setup. 
In Sec.~(\ref{sec:results}), we report the measurement results obtained with commercial detectors and with the GPD under thermal vacuum (TV) conditions.
We discuss the results in Sec.~(\ref{sec:discussion}) and present our conclusions in Sec.~(\ref{sec:conclusion}).

\section{Methods}
\label{sec:methods}
We present the FCW and the four polarized and unpolarized calibration sources contained therein (Sec. \ref{sec:fcs}). 
Subsequently, in Sec.~(\ref{sec:setup}), we describe the experimental setup: first, the setup under clean room conditions with a commercial silicon drift detector (SDD) and charge-coupled device (CCD), followed by the setup  under TV conditions with the FMs of the GPD.
In Sec.~(\ref{sec:swtools}), we describe the data analysis tools.
\subsection{Filter and Calibration Set}
\label{sec:fcs}
\begin{figure}
\begin{center}
\begin{tabular}{c}
\includegraphics[height=5.5cm]{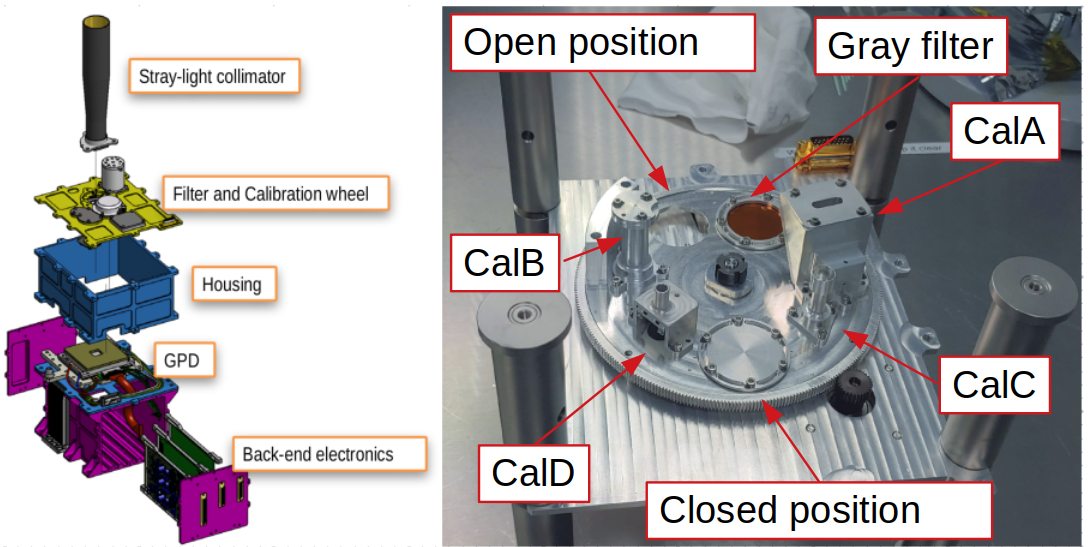}   
\\
(a) \hspace{5.1cm} (b)
\end{tabular}
\end{center}
\caption 
{ \label{fig:duexploded}
(a) Deconstructed diagram of IXPE DU, showing the position of the FCW on top of the GPD; 
(b) picture of the FCS installed onto the wheel in the IAPS clean room.} 
\end{figure}

The FCS is composed of four calibration sources (a polarized source, CalA; a collimated unpolarized source, CalB; two uncollimated, unpolarized sources, CalC and CalD), a gray filter, and an open and closed position. 
The FCS is hosted in the FCW, as shown in Fig.~(\ref{fig:duexploded}) (b), which is placed on the top lid of the DU, as shown in Fig.~(\ref{fig:duexploded}) (a).
By rotating around its central axis, the FCW allows the placement of one of the four calibration sources or a gray filter in front of the GPD, in addition to the open and closed position, depending on the observational requirements.
The FCW also hosts other elements that ensure the stability and positional accuracy of the calibration sources.
A rotary potentiometer is used to determine the wheel angular position accurately. 
Further, for redundancy, three radially placed Hall effect sensors and twelve magnets (positioned to realize a unique binary coding for the wheel's seven positions) function as position reference points.
The calibration sources can thus be positioned with an accuracy greater than $\pm$500 $\mu$m with respect to their nominal positions.
The angular position of the polarized calibration source with respect to the DU coordinate system is known with an uncertainty below 20 arcmin.
The fixed parts of the FCW (e.g., the cover lid) are connected to the rotating parts (the wheel itself) by a bearing sub-assembly.
Finally, a ballast mass is installed to balance the weights and the momentum of inertia on the wheel.
Each calibration source inside the FCW contains a radioactive source constituted of a $^{55}$Fe nuclide, which, following a K electron capture, emits X-rays at 5.9 and 6.5 keV, i.e., the Mn K$\alpha$ and Mn K$\beta$ emission lines, respectively.
The activity of $^{55}$Fe naturally decays with a half-life of 2.7 years, which provides sufficient time to cover the entire operative life of IXPE.
This solution removes the problem of including X-ray tubes on board of the spacecraft, as their installation would have been complex, as well as mass- and power-demanding, especially on a moving support. \\
In Table~(\ref{tab:requirements}), we list the rate requirements for each source that have to be satisfied by the end of the tests presented in this paper, as well as the scientific observable that can be obtained from each source. 
The requirements regarding the counting rate are set by the statistical significance required to validate the results and by the activity of the radioactive sources on board.
A key quantity in X-ray polarimetry is indeed the minimum detectable polarization (MDP)\cite{2010Weisskopf}, which is defined as the minimum polarization degree that can be detected with certain confidence, given the source intensity and observation time.
The MDP also depends on detector characteristics, which are represented by the modulation factor $\mu$.
Given that the modulation factor is itself a characteristic that must be determined by calibration, a measurable quantity is the minimum detectable amplitude (MDA)\cite{2010Weisskopf}, which depends only on the number of detected counts $N$:
%
\begin{equation}
MDA = \frac{4.29}{\sqrt{N}} \quad.
\end{equation}
Given a certain rate $R=N/T$, where $T$ is the duration of the measurement, the time needed to reach a 1\% MDA is
\begin{equation}
T = \Big(\frac{4.29}{MDA_{1\%}}\Big)^2 \frac{1}{R} \simeq 184\times10^3 \frac{1}{R} \quad.
\end{equation}
A counting rate of at least $R>3$ c/s assures that the counts needed to reach an MDA of at least 1\% can be collected in less than a day. 
A certain number of counts for an MDA/2 must be collected to achieve a 1$\sigma$ measurement of a certain amplitude\cite{2013Strohmayer}.
Thus, the MDA is not the uncertainty but an indicator of the confidence of the measurement.
The ultimate purpose is to evaluate the sensitivity to polarization, such that the measured MDA is divided by the modulation factor $\mu$ to determine the minimum observable polarization and the absolute error on the measurement of the polarization degree.
Other important observables are the photoelectron track size and length: the track size is defined by the number of contiguous pixels, in which the charge is acquired for a single event, and the track length is defined as the second momentum of the track.
Their correct determination allows us to distinguish between noise and real events, determine the polarization direction of the latter, and check the pressure of the gas in the GPD. \\
In the following sections, we describe the four calibration sources in detail.
\begin{table}[htbp]\footnotesize
\caption{Spatial and rate requirements and scientific observables are reported for each on-board calibration source.} 
\label{tab:requirements}
\begin{center}       
\begin{tabular}{|l|l|l|} 
\hline
\rule[-1ex]{0pt}{3.5ex}  Calibration 			& Requirements 										& Scientific   \\
\rule[-1ex]{0pt}{3.5ex}  Source 				& 			 										& observables  \\
\hline\hline
\rule[-1ex]{0pt}{3.5ex}  \multirow{3}{*}{Cal A}	& Polarized beam illuminating entire detector;	& Modulation factor and energy resolution   \\
\rule[-1ex]{0pt}{3.5ex}							& rate $>$3 c/s at 3 keV;							& at 3 and 5.9 keV, counting rate,  \\
\rule[-1ex]{0pt}{3.5ex}							& rate $>$40 c/s at 5.9 keV 						& track length and track size, gain  \\
\hline
\rule[-1ex]{0pt}{3.5ex}  \multirow{2}{*}{Cal B}	& Collimated 3 mm beam at the center of GPD;		& Spurious modulation, energy resolution \\
\rule[-1ex]{0pt}{3.5ex}							& rate $>$30 c/s at 5.9 keV							& at 5.9 keV, counting rate, gain  \\
\hline 
\rule[-1ex]{0pt}{3.5ex}  \multirow{3}{*}{Cal C}	& Uncollimated beam illuminating entire 			& Gain, counting rate, energy resolution    \\
\rule[-1ex]{0pt}{3.5ex}							& 15$\times$15 mm$^2$ surface of GPD;			& at 5.9 keV, spurious modulation  \\
\rule[-1ex]{0pt}{3.5ex}							& rate $>$80 c/s at 5.9 keV 						& 			  \\
\hline 
\rule[-1ex]{0pt}{3.5ex}  \multirow{3}{*}{Cal D}	& Uncollimated beam illuminating entire 			& Gain, counting rate, energy resolution  \\
\rule[-1ex]{0pt}{3.5ex}							& 15$\times$15 mm$^2$ surface of the GPD;			& at 1.7 keV, spurious modulation  \\
\rule[-1ex]{0pt}{3.5ex}							& rate $>$10 c/s at 1.7 keV 						&   \\
\hline 
\end{tabular}
\end{center}
\end{table} 
\subsubsection{Calibration source A (CalA)}
\begin{figure}[htbp]
\begin{center}
\begin{tabular}{c}
\includegraphics[height=5.5cm]{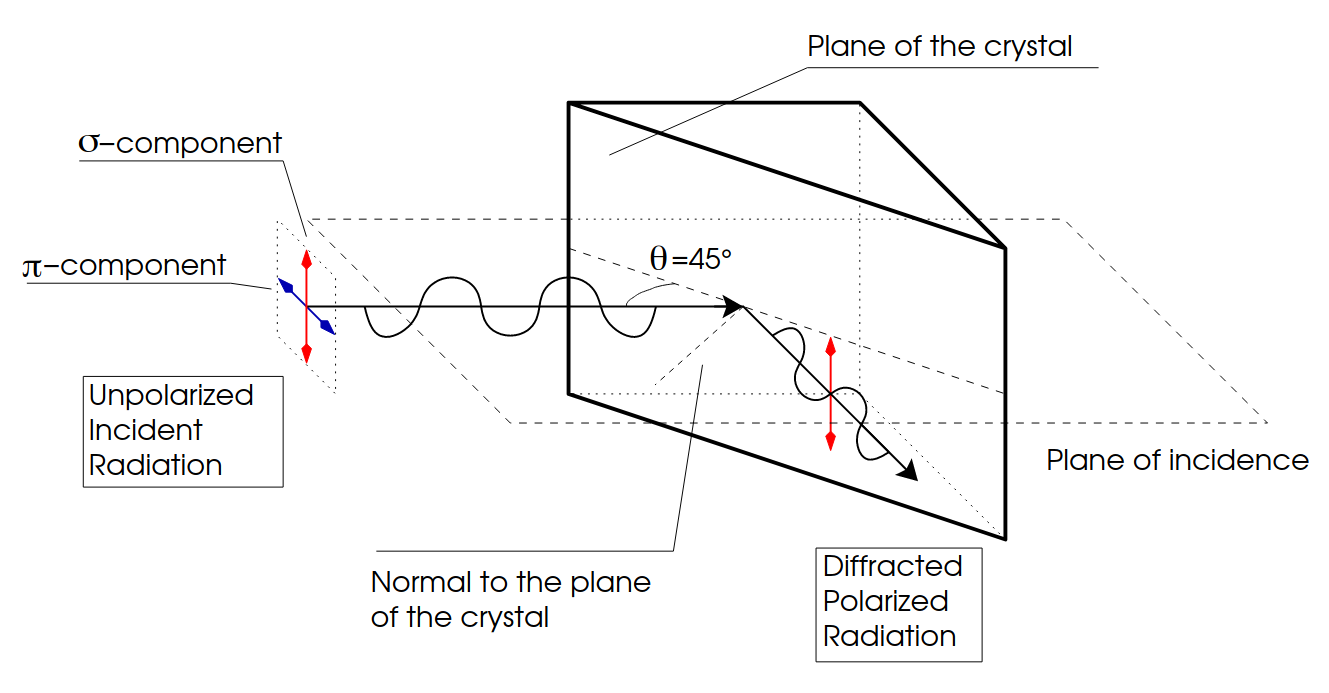}
\end{tabular}
\end{center}
\caption 
{ \label{fig:diffrazionebragg}
Bragg diffraction on a crystal at $45^\circ$ (Reproduced with permission from\cite{2008Muleri}). 
The radiation component polarized perpendicularly to the plane of diffraction is efficiently diffracted, while the radiation component that has parallel polarization with respect to such a plane is absorbed by the crystal. } 
\end{figure}

\begin{figure}[htbp]
\begin{center}
\begin{tabular}{c}
\includegraphics[height=5.5cm]{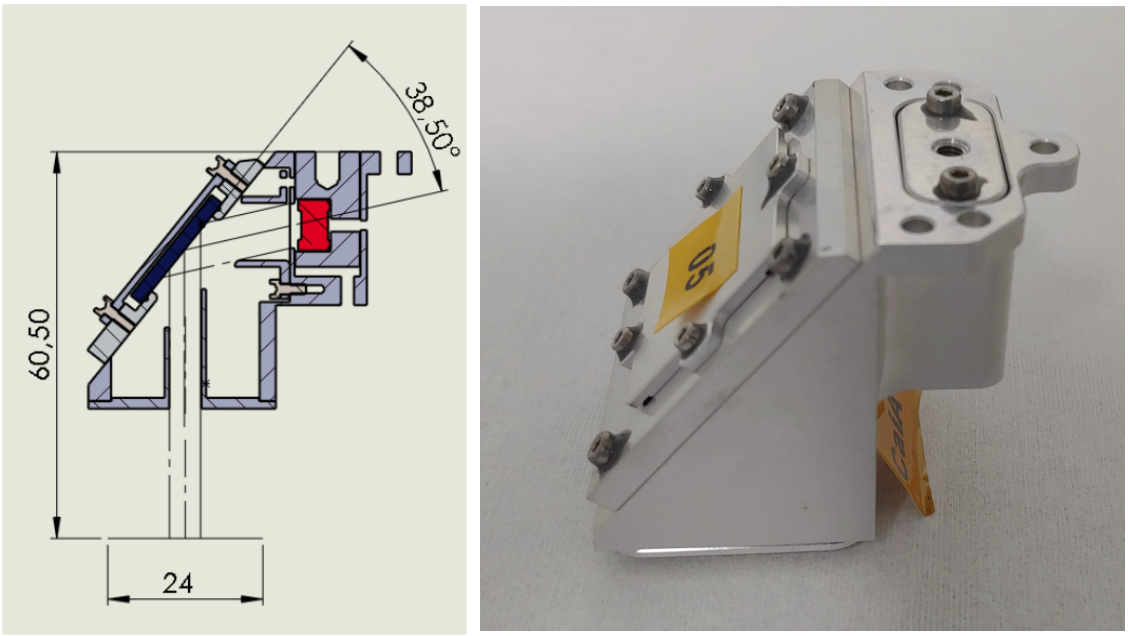}  
\\
(a) \hspace{5.1cm} (b)
\end{tabular}
\end{center}
\caption 
{ \label{fig:cala}
The polarized on-board calibration source Cal A enables production of polarized X-ray photons of known energy.
(a) Cross-sectional view of CalA: the $^{55}$Fe radioactive source is shown in red; a thin silver foil is placed in front to extract 3.0 keV fluorescence. The graphite mosaic crystal for Bragg diffraction is shown in dark blue. Numbers are in mm.
(b) Picture of CalA.} 
\end{figure}

\begin{figure}[htbp]
\begin{center}
\begin{tabular}{c}
\includegraphics[height=8cm]{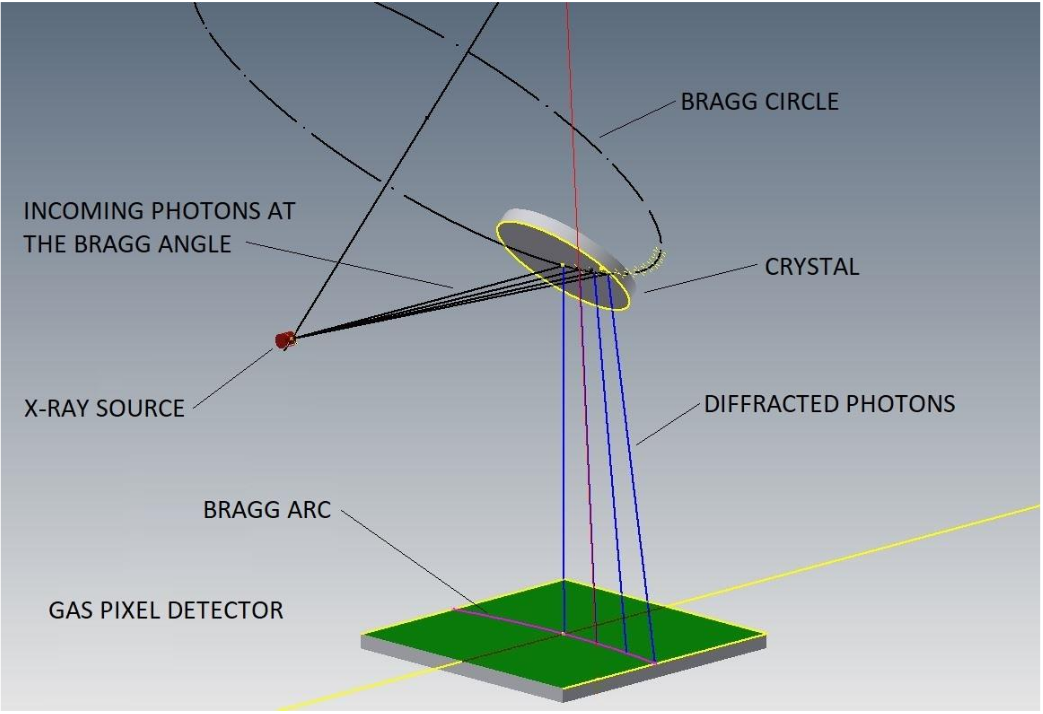}
\end{tabular}
\end{center}
\caption 
{ \label{fig:braggarccad}
	Geometry of Bragg diffraction for monochromatic photons. } 
\end{figure}

CalA \cite{2007Muleri,2008Muleri}, shown in Fig.~(\ref{fig:cala}), generates polarized X-ray photons of a precisely-known energy and polarization state, allowing for the monitoring of the modulation factor of the instrument at two energies, 3 and 5.9 keV, in the IXPE energy band (2 - 8 keV).
Its working principle is based on Bragg diffraction, defined as the superposition of coherent Thomson scatterings on a periodic medium, such as the lattice of a crystal.
Bragg diffraction is effective for energies $<$10 keV, and only when the Bragg law is satisfied:
\begin{equation}
E = \frac{nhc}{2d\sin{\theta}} \, ,
\label{eq:Condizione_Bragg}
\end{equation}
where $\theta$ is the grazing angle of incidence measured with respect to the crystal surface, as shown in  Fig.~( \ref{fig:diffrazionebragg}); $E$ is the photon energy; $d$ is the crystal lattice step; $n$ is the diffraction order; $h$ is the Planck constant; and $c$ is the speed of light. 
As shown in Figure~(\ref{fig:diffrazionebragg}), unpolarized incident radiation that satisfies Eq.~(\ref{eq:Condizione_Bragg}), can be decomposed in two orthogonally polarized components: the $\pi$-component, in which the polarization lies parallel to the plane of incidence (defined as the plane of the direction of incident radiation and the normal of diffracting planes), and the $\sigma$-component is normal to this plane. 
While the former is absorbed by the crystal, the latter is diffracted; therefore, Bragg diffraction at around 45$^\circ$ polarizes the sources completely. 
This process has a narrow energy band ($<$1 eV for a perfect crystal), and its efficiency decreases rapidly as the photon wavelength or incidence angle changes.
%
This major limitation can be partially overcome by employing mosaic crystals, i.e., composed of small domains, each acting as an independent crystal. Because the domains are smaller than the absorption length of X-ray photons, radiation crosses numerous domains and can be diffracted, provided that one of the domains is properly aligned, at angles (i.e., energies) that are slightly different from the average ones.
This increases the energy bandwidth to several tens of eV, slightly reducing the mean degree of polarization of diffracted radiation, for a continuum spectrum only. 
The initial design of CalA was based on 5.9 and 6.5 keV photons emitted by a $^{55}$Fe nuclide, partially absorbed by a thin polyvinyl chloride (PVC) film to extract the 2.7 keV Chlorine fluorescence line.
The 2.7 and 5.9 keV photons were then diffracted at first order at nearly 45$^{\circ}$ by a graphite crystal and a lithium-fluoride crystal, respectively. 
Because regulations on the use of materials for space applications do not recommend the use of PVC due to outgassing, a different source for the second polarized energy line had to be found. 
Eventually, we adopted a design in which X-rays from the $^{55}$Fe nuclide at 5.9 and 6.5 keV are partially absorbed by a 1.6 $\mu$m thick silver foil to produce L$\alpha1,\alpha2$ fluorescence lines at 2.99 keV and an L$\beta$ line at 3.15 keV. 
The silver foil is deposited between two polyimide foils of 8 $\mu$m thickness (on the side towards the $^{55}$Fe) and 2 $\mu$m thickness (on the side of the crystal) with negligible X-ray losses. 
Photons at $\sim$3.0 and 5.9 keV, collimated with a broad collimator, are subsequently Bragg-diffracted on a graphite mosaic crystal, with full width half maximum mosaicity of 1.2$^{\circ}$, at the first and second orders of diffraction, approximately at the same diffraction angle (38.3$^{\circ}$ and 38.7$^{\circ}$, respectively) and hence with the same polarization degree ($\sim$67\% \cite{1993Henke}). 
A second broad collimator is used to block stray-light X-rays. \\
%
Given a point-like source, the position of points on the plane of the crystal satisfying the Bragg condition is a circle, and its projection on the detector will appear as an arc (see Fig.~(\ref{fig:braggarccad})).
We call this the "Bragg arc", whose width depends on the X-ray source and crystal used.
Polarization along the arc is expected to remain aligned with the polarization angle tangent to the Bragg arc.
In the case of CalA, the image of the diffracted photons appears as a 4 mm wide, slightly curved, strip extending across the detector (see for example Fig.~(\ref{fig:ccd_FM02})). 
This can be utilized to study the response of the entire detector to polarized radiation by moving the wheel.  \\
We named the four sources CalA1, CalA2, CalA3, and CalA4 according to the FCS they are part of.
%
\subsubsection{Calibration source B (CalB)}
\begin{figure}[htbp]
\begin{center}
\begin{tabular}{c}
\includegraphics[height=5.5cm]{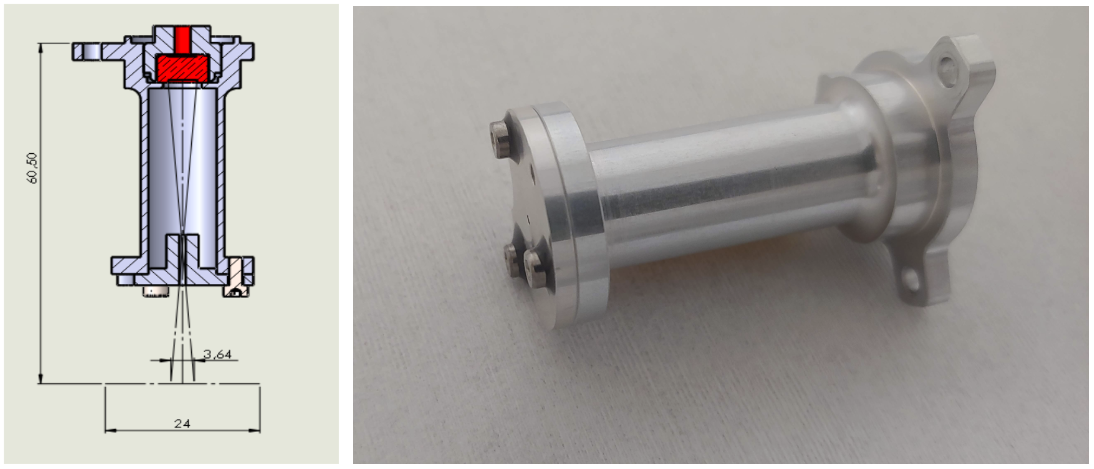}  
\\
(a) \hspace{5.1cm} (b)
\end{tabular}
\end{center}
\caption 
{ \label{fig:calb}
CalB illuminates the GPD with a monochromatic unpolarized X-ray, simulating the region illuminated by a point-like source.
(a) Cross-sectional view of CalB: the $^{55}$Fe radioactive source is shown in red. Numbers are in mm.
b) Picture of CalB.} 
\end{figure}

The CalB source produces a collimated beam of unpolarized photons to monitor the absence of a spurious modulation.  
A $^{55}$Fe radioactive source is glued on a holder and screwed in a cylindrical body, at the end of which a diaphragm with an aperture of 1 mm collimates X-rays to produce a spot of about 3 mm diameter on the GPD. 
This spot has a size that is representative of the region illuminated by the photons of a point-like source when the spacecraft pointing dithering strategy is actuated. 
A cross-sectional diagram and a photograph of the source are shown in Fig.~(\ref{fig:calb}). \\
We named the four sources CalB1, CalB2, CalB3, and CalB4 according to the FCS they are part of.
%
\subsubsection{Calibration source C (CalC)}
\begin{figure}[htbp]
\begin{center}
\begin{tabular}{c}
\includegraphics[height=5.5cm]{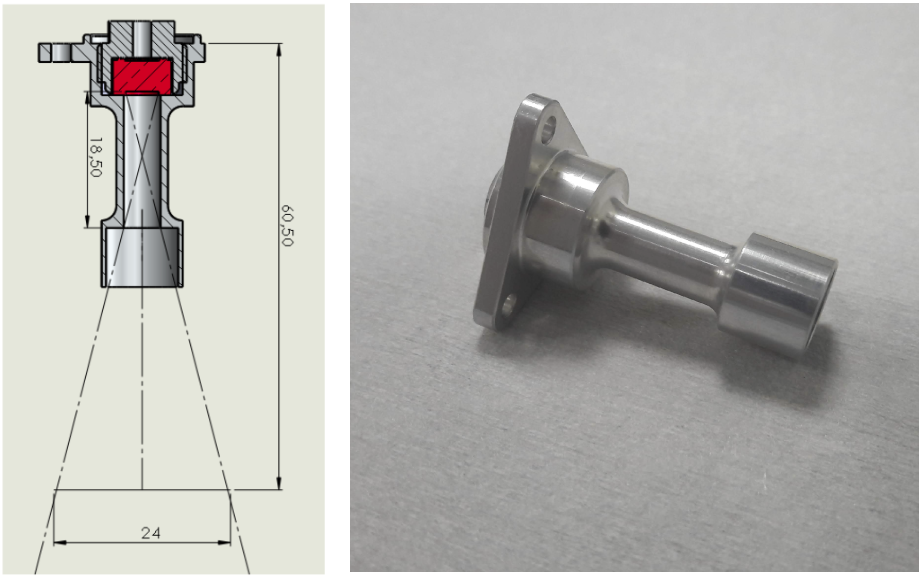} 
\\
(a) \hspace{5.1cm} (b)
\end{tabular}
\end{center}
\caption 
{ \label{fig:calc}
CalC illuminates the entire GPD surface with monochromatic unpolarized photons.
(a) Cross-sectional view of CalC: the $^{55}$Fe radioactive source is shown in red. Numbers are in mm. 
(b) Picture of CalC.} 
\end{figure}

This source illuminates the entire detector sensitive area at a certain energy (5.9 keV) to map the changes in the gain (i.e., the relation between the pulse height (PHA), which is proportional to the charge collected by the detector and the energy of the photons), as a function of the position.
This source is a $^{55}$Fe radioactive source, glued in a holder similar to CalB, but with a diaphragm-less collimator that allows X-ray photons to impinge on the whole detector sensitive area.
A cross-sectional diagram and a photograph of the source are shown in Fig.~(\ref{fig:calc}).  \\
We name the four sources CalC1, CalC2, CalC3, and CalC4 according to the FCS they are part of.
%
\subsubsection{Calibration source D (CalD)}
\begin{figure}[htbp]
\begin{center}
\begin{tabular}{c}
\includegraphics[height=5.5cm]{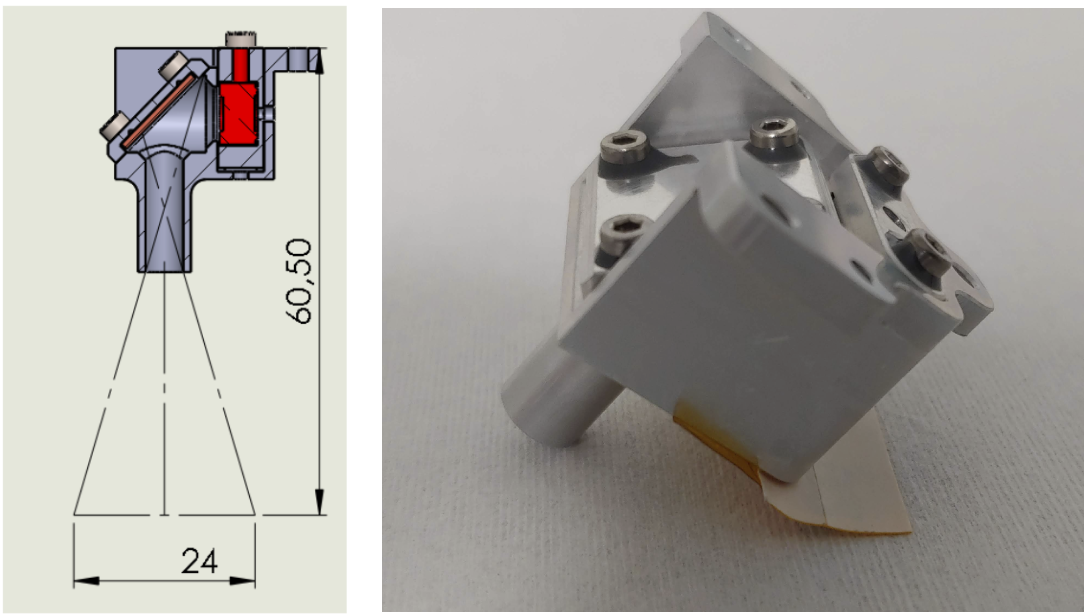}  
\\
(a) \hspace{5.1cm} (b)
\end{tabular}
\end{center}
\caption 
{ \label{fig:cald}
CalD illuminates the whole GPD surface with monochromatic unpolarized photons with lower energy than CalC.
(a) Cross-sectional diagram of CalD: the $^{55}$Fe radioactive source is shown in red; the silicon target for 1.7 keV fluorescence is shown in orange.
(b) Picture of CalD.} 
\end{figure}

This source illuminates the entire detector sensitive area as CalC, to map the gain on the entire GPD surface at a different energy. 
CalD is based on a $^{55}$Fe source, which illuminates a silicon target to extract K fluorescence from silicon at 1.7 keV, which impinges on the detector. 
Because the length of the photoelectron tracks is a function of the impinging photon energy, at 1.7 keV the CalD tracks are shorter than the 5.9 keV track from CalC.
This allows us to map the gain of the GPD at a higher spatial resolution. 
Moreover, the large energy difference between the 1.7 keV energy of CalD and 5.9 keV energy from CalC allows us to determine the calibration relation (PHA/Energy) with high accuracy.
CalD is designed such that X-ray photons from $^{55}$Fe cannot directly impinge on the GPD sensitive area, avoiding detector saturation. 
Some of these photons are scattered by silicon and impinge on the detectors with an energy that is almost unchanged. 
A cross-sectional diagram and a photograph of the source are shown in Fig.~(\ref{fig:cald}). \\
We name the four sources CalD1, CalD2, CalD3, and CalD4 according to the FCS they are part of.
%
\subsubsection{Open position, closed position, gray filter}
In the FCS, apart from the four calibration sources, there is also an open position, a closed position, and a gray filter. 
The open position  is the standard science operation position.
The closed position (i.e., a black filter) is a lid opaque to X-ray radiation (with transparency lower than 10$^{-6}$ at 8 keV) used to cover the detector during the internal background measurements.
The gray filter will be employed when observing very bright sources (with a flux higher than $\sim$4$\times$ 10$^{-8}$ erg/cm$^2$ /s, or two Crab in the 2 - 8 keV range).
It received its name in analogy with filters that have the same function in the optical band; however, in contrast to those, the transparency of the IXPE is strongly energy dependent. 
The opacity of the gray filter is such that the flux for a source with a power law with spectral index of two is reduced by a factor of eight in the 1 - 12 keV energy range, implying a transparency of 14\% at 2.6 keV.
However, this does not affect the response to polarization: while being more opaque to softer X-ray photons, which are typically those with a lower polarization, it will improve the statistics at higher energies, which are of major astrophysical interest.
\subsection{Experimental setup}
{\label{sec:setup}}
The experimental setup was designed to study the performance of the FCS prior to its installation.
To this aim, we used a SDD for the spectrum and a CCD to acquire images. 
Because no commercial standard X-ray polarimeter exists, the polarimetric performances are tested with the GPD. \\
In Table~(\ref{tab:radioactivesources}) we report the activity and emission rate of the $^{55}$Fe radioactive sources employed during the measurements with the SDD and in the TV chamber with the DU.
The sources are manufactured by Eckert \& Ziegler, providing a nominal activity and a measured emission rate for each nuclide: while the former can be affected by errors as high as 30\% due to self-absorption, the latter can be measured with greater accuracy, hence representing a better estimate of the $^{55}$Fe source activity.
Therefore, we use the emission rate to calculate the expected flight rate.\\
For CalC, which directly illuminates the detector, a weaker source can be employed.
When integrated on the spacecraft, the calibration sources will be equipped with radioactive sources of higher nominal activity: 100 mCi for CalA, 20 mCi for CalB, 0.5 mCi for CalC, and 100 mCi for CalD.
We refer to these as "flight radioactive sources".
The emission rate of the flight radioactive sources has been measured, and the results are listed in Table~(\ref{tab:radioactivesources}).
Due to the high radioactivity of some of these sources, tests on them could not be performed in the laboratory environment. 
The weaker $^{55}$Fe nuclides already present in our lab were used in the tests, and the flight rate (at the beginning of the mission) is inferred by multiplying the measured rate by the ratio between the emission rate of the flight sources at the beginning of the mission operations and that of our sources at the time of measurement. 
In the following, we refer to the "expected flight rate" as the rate of the calibration sources using the flight radioactive sources.
To maximize the activity of the on-board nuclides to achieve the required counting rate, flight radioactive sources will be installed in the instrument as late as possible in the integration flow. 
%
\begin{table}\tiny
\caption{$^{55}$Fe radioactive nominal source activity and emission rate at the time of tests with SDD and GPD.
		The date at which the nominal activity was measured is given in parentheses.
		The last row lists the activity of the flight radioactive sources and the emission rate of the nuclides.
		The emission rate of the source employed in CalC measurements with SDD is not known.}
\label{tab:radioactivesources}       
\begin{center}  
\begin{tabular}{|l|l|l|l|l|l|l|l|l|}
\hline 
\rule[-1ex]{0pt}{3.5ex}	 	& \multicolumn{2}{c|}{CalA}	& \multicolumn{2}{c|}{CalB} & \multicolumn{2}{c|}{CalC} & \multicolumn{2}{c|}{CalD} \\
\hline
\rule[-1ex]{0pt}{3.5ex}\multirow{3}{*}{Measurement} & Activity 	& Emission 				& Activity 	& Emission  			& Activity 	& Emission  & Activity & Emission  \\
\rule[-1ex]{0pt}{3.5ex}								& [mCi]  	& rate					& [mCi]  	& rate 					& [mCi]  	&  rate 	& [mCi]  &  rate \\ 
\rule[-1ex]{0pt}{3.5ex}								& (date) 	& [s$^{-1}$strd$^{-1}$] & (date)	& [s$^{-1}$strd$^{-1}$]	& (date)	& [s$^{-1}$strd$^{-1}$] & (date)& [s$^{-1}$strd$^{-1}$] \\
\hline\hline
\rule[-1ex]{0pt}{3.5ex}Nominal			 	& 8.1 (2017/01/16)	& 6.55E6	& 8.1 (2017/01/16)	& 6.55E6		& 5 (2018/01/18)&	-					& 8.1 (2017/01/16)& 6.55E6	\\ 
\hline
\rule[-1ex]{0pt}{3.5ex}SDD FM1 				& 5.13 	& 2.81E6 	& 5.13 	& 2.81E6		& 0.34 		& 	-					& 5.13 		& 2.81E6 \\	
\rule[-1ex]{0pt}{3.5ex}SDD FM2,3,4 			& 4.82 	& 2.42E6 	& 4.82 	& 2.42E6		& 0.31 		& 	-					& 4.82 		& 2.42E6 \\
\hline\hline	
\rule[-1ex]{0pt}{3.5ex}Nominal				& 8.1 (2017/01/16)	& 6.55E6	& 0.50 (2019/07/01)	& 2.32E5		& 0.50 (2019/07/01) & 2.18E5				& 4	(2019/07/01)	& 2.01E6		\\ 
\hline
\rule[-1ex]{0pt}{3.5ex}DU1 with FCW1 in TV	& 4.22 	& 2.11E6	& 0.49 	& 2.29E5		& 0.49 		& 2.15E5				& 3.95 		& 2.01E6 \\	
\rule[-1ex]{0pt}{3.5ex}DU2 with FCW2 in TV	& 4.09 	& 2.06E6	& 0.48	& 2.22E5		& 0.48 		& 2.09E5				& 3.83 		& 1.93E6 \\	
\rule[-1ex]{0pt}{3.5ex}DU3 with FCW3 in TV	& 4.03 	& 2.02E6	& 0.47	& 2.19E5		& 0.47 		& 2.06E5				& 3.78 		& 1.90E6 \\	
\rule[-1ex]{0pt}{3.5ex}DU4 with FCW4 in TV	& 3.96 	& 1.98E6	& 0.44	& 2.02E5		& 0.44		& 2.15E5				& 3.71 		& 1.86E6 \\
\hline	
\rule[-1ex]{0pt}{3.5ex}	\multirow{4}{*}{Flight radioactive sources}	
			& \multirow{2}{*}{100}	& 2.81E7 (DU1)& \multirow{2}{*}{20}	& 7.10E6 (DU1)	&\multirow{2}{*}{0.5}	& 1.53E5 (DU1)& \multirow{2}{*}{100}& 2.82E7 (DU1)\\
			& 						& 2.82E7 (DU2)&						& 7.54E6 (DU2)	&						& 1.70E5 (DU2)&						& 2.85E7 (DU2)\\
			&\multirow{2}{*}{(2020/04/29)} & 3.16E7 (DU3)& \multirow{2}{*}{(2020/04/29)}& 8.14E6 (DU3)	&\multirow{2}{*}{(2019/12/18)}& 2.07E5 (DU3)&\multirow{2}{*}{(2020/04/29)}	& 3.09E7 (DU3)\\
			& 						& 3.06E7 (DU4)&					& 7.64E6 (DU4)	&						& 2.06E5 (DU4)&	 & 3.06E7 (DU4)\\
\hline 
\end{tabular}
\end{center}
\end{table}
\subsubsection{Calibration source spectra and images with SDD and CCD}
\begin{figure}
	\begin{center}
		\begin{tabular}{c}
			\includegraphics[height=4.45cm]{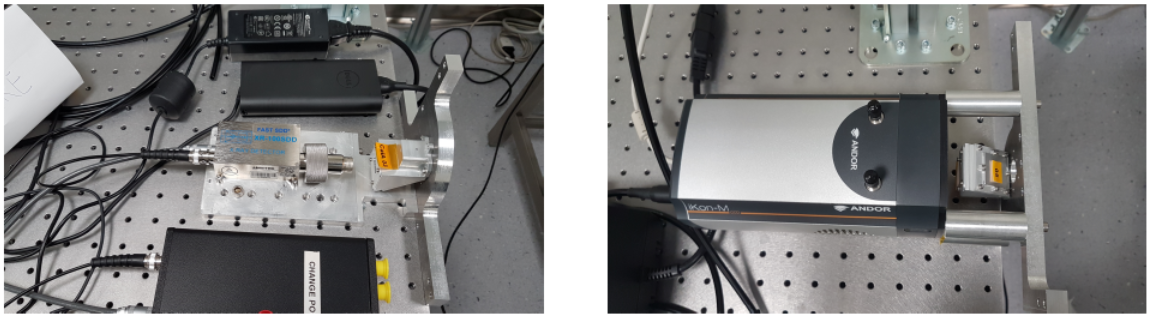}
			\\
			(a) \hspace{8.1cm} (b)
		\end{tabular}
	\end{center}
	\caption 
	{ \label{fig:experimentalsetup}
		(a) CalA in front of SDD for spectrum measurement.
		(b) CalA in front of CCD for image acquisition.} 
\end{figure}

The spectrum of each source is acquired with an Amptek XR-100SDD SDD (see Fig.~(\ref{fig:experimentalsetup}) (a)) to verify that X-rays are correctly emitted. 
The tests are performed inside an ISO 7 clean room with controlled temperature and humidity conditions.
The measurement of spectra in air is affected by the absorption of X-rays, with the transmission factor $\tau(E)$ given by the relation:
%
\begin{equation}
\tau(E) = \frac{I}{I_0} = e^{-\mu_{air}(E)\rho d} \,
\label{eq:transmisison}
\end{equation}
where $I/I_0$ is the fraction of photons arriving on the detector, $\mu_{air}$ is the energy-dependent total mass attenuation coefficient of air (in cm$^2$/g) adopted from online tables of the National Institute of Standards and Technology\cite{NIST}, $\rho = 1.225\times10^{-3}$ g/cm$^3$ is the mass density of air, and $d$ is the distance between the source and detector.
Low energy sources, like the Ag line of CalA and the Si line from CalD, are the most affected by air absorption.
Moreover, the argon in the air, excited by the X-rays emitted by the $^{55}$Fe inside the sources, emits at an energy of $2.96$ keV, which is very similar to the L line of silver, thus contaminating the Cal A measurements.
Even if the fluorescence yield of Ar is moderate (12.3\% \cite{1972Bambynek}) and only a limited fraction of the fluorescence photons is emitted toward the detector, this distortion must be accounted for.
Knowing the distance between source and detector, fixed by the mechanical setup, as well as the air properties, we can derive the losses due to X-ray absorption in air.
As for the argon contamination, rather than performing complex calculations strongly dependent on geometric details, we solved this problem by placing the DU, CalA included, in the chamber for TV tests, as described later. 
The counting rate measured with the SDD is then used to extrapolate the expected GPD rate through the relation
%
\begin{equation}
R_{GPD} = \sum^E R_{SDD}(E) \epsilon(E) \frac{1}{\tau(E)} \Big(\frac{A_{GPD}}{A_{SDD}}\Big) \Big(\frac{r_{flight}}{r_0}\Big)  \,
\label{eq:sddtogpdrate}
\end{equation}
where $\epsilon(E)$ is the GPD efficiency at energy $E$, $\tau(E)$ is the transmission factor as defined in Eq.~( \ref{eq:transmisison}), $A_{GPD}$ is the area illuminated by the source on the GPD, and $A_{SDD}$ the area illuminated on the SDD. 
The area ratio accounts for the loss of photons due to differences in the detector area, especially for CalC and CalD, as the 15$\times$15 mm$^2$ area of the GPD is larger than the 7$\times$7 mm$^2$ area of the GPD. 
For CalB, the area ratio is 1, as the beam is collimated, and there are no spatial losses. In contrast, for CalA, we have to account for the shape of the Bragg arc, which can be approximated with a 14$\times$6 mm$^2$ rectangle on GPD, whereas on the SDD, a 7$\times$6 mm$^2$ region is assumed to be illuminated.
The energies are summed, because the SDD has an higher spectral resolution than the GPD, and it is able to detect both $K_\alpha$ and $K_\beta$ emission lines, which are not resolved by the GPD and hence considered together. 
Finally, $r_{flight}$ is the emission rate of the $^{55}$Fe radioactive source at the start of the mission, as reported in Tab.~(\ref{tab:radioactivesources}), while $r_0$ is the emission rate at the time of the measurement. 
By multiplying by their ratio, the expected counting rate of each source at the beginning of the mission can be estimated.
Certainly, due to the corrections needed and the contamination by the argon K$\alpha$ line in air, the measurements taken during TV tests provide a better estimate of the flight rate.
We recall here that we assumed the nominal activity at launch, while the real activity will be determined only at a later time. 
As for the source employed during the SDD tests of CalC, the emission rate is unknown, we employ the nominal activity to estimate the flight rate.
In Table~(\ref{tab:commonsetup}), for each calibration source, we list the experimental parameters common to all measurements with the SDD: the area illuminated on the SDD and the GPD, the air path, i.e., the linear distance between the source and the SDD, the energy of each line detected by the SDD, the GPD efficiency at that energy, and the air mass-attenuation coefficient at that energy.
%
\begin{table}[htbp]\footnotesize
\caption{Parameters common to SDD measurements of all FMs. Energy lines were adopted from \cite{Xraydatabooklet}; mass attenuation coefficients were adopted from \cite{NIST}.}
\label{tab:commonsetup}       
\begin{center}
\begin{tabular}{|l|l|l|l|l|l|l|}
\hline
\rule[-1ex]{0pt}{3.5ex}\multirow{4}{*}{Source}	& Area on 	& Area on  	& Air	& Energy (Line)		& GPD 			& Mass attenuation\\
\rule[-1ex]{0pt}{3.5ex}							& SSD		& GPD		& path	& [kev]				& efficiency	& coefficient\\		
\rule[-1ex]{0pt}{3.5ex}							& $A_{SDD}$	& $A_{GPD}$	& $d$	& 					& $\epsilon$	& $\mu_{air}$	\\	
\rule[-1ex]{0pt}{3.5ex}							& [mm$^2$]	&[mm$^2$]	& [cm]	& 					& 				& [cm$^2$/g]\\	
\hline\hline
\rule[-1ex]{0pt}{3.5ex}\multirow{4}{*}{CalA}	& \multirow{4}{*}{42.0}& \multirow{4}{*}{84.0} & \multirow{4}{*}{4.88}	& 2.9 (Ag L$\alpha$)	& 0.150 & 165.3\\ 
\rule[-1ex]{0pt}{3.5ex}							& 			&			&		& 3.1 Ag (L$\beta$) & 0.134 & 140.4 \\
\rule[-1ex]{0pt}{3.5ex}							& 			&			&		& 5.9 (Mn K$\alpha$)& 0.025 & 41.74 \\
\rule[-1ex]{0pt}{3.5ex}							& 			&			&		& 6.5 (Mn K$\beta$)	& 0.018 & 31.64 \\
\hline
\rule[-1ex]{0pt}{3.5ex}		\multirow{2}{*}{CalB}& \multirow{2}{*}{7.1}& \multirow{2}{*}{7.1} &\multirow{2}{*}{5.39}	& 5.9 (Mn K$\alpha$) & 0.025 & 41.74 \\ 
\rule[-1ex]{0pt}{3.5ex}							& 			& 			&		& 6.5 (Mn K$\beta$)	& 0.018 & 31.64 \\
\hline
\rule[-1ex]{0pt}{3.5ex}		\multirow{2}{*}{CalC}& \multirow{2}{*}{49.0}& \multirow{2}{*}{225.0}& \multirow{2}{*}{5.39} & 5.9 (Mn K$\alpha$) & 0.025 & 41.74 \\ 
\rule[-1ex]{0pt}{3.5ex}							& 			&			& 		& 6.5 (Mn K$\beta$)	& 0.018 & 31.64\\
\hline
\rule[-1ex]{0pt}{3.5ex}		CalD				& 49.0		& 225.0		& 5.31 	& 1.7 (Si K$\alpha$)& 0.217 & 777.8 \\
\hline
\end{tabular}
\end{center}
\end{table}
%
The image of the X-rays emitted from the sources is acquired with an Andor iKon-M 934 CCD camera (see Fig.~( \ref{fig:experimentalsetup}) (b)).
Because the CCD surface (13$\times$13 mm$^2$) is similar to the GPD surface (15$\times$15 mm$^2$), their images are similar. 
\subsubsection{Thermal vacuum measurement with GPD }
\label{sec:termovacuum}
\begin{figure}
\begin{center}
\begin{tabular}{c}
\includegraphics[height=8.5cm]{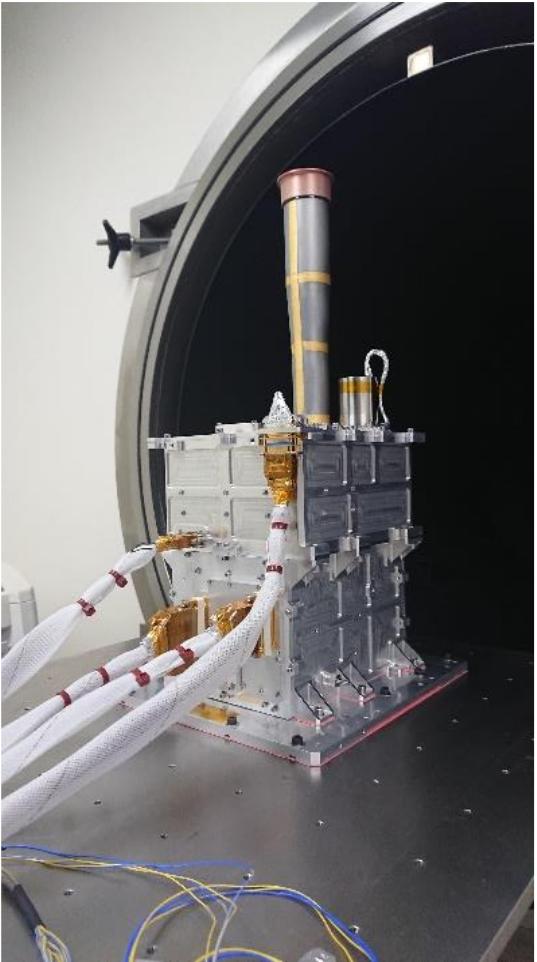}
\end{tabular}
\end{center}
\caption 
{ \label{fig:dutv}
DU with FCS4 in front of the TV chamber before the tests.} 
\end{figure}

After the tests with SDD and CCD, the calibration sources were installed in the FCWs and integrated in the three DU FMs and the spare unit.
We subsequently tested them in the INAF-IAPS TV facility to measure a more accurate expected rate at the launch, a detailed response of the DU to the sources installed in the FCW and the centering of each source.
During the tests, the temperature was constant and the pressure inside the TV chamber was about 1E-6 mbar. This pressure is significantly smaller than 0.01 mbar, which is the maximum pressure allowed before discharges occur in the DU, as derived using the Paschen law while considering a breakdown voltage of 3 kV and a maximum distance of 2 cm to grounded elements. The breakdown voltage and maximum distance are chosen by considering the high voltage board design. 
We monitored the vacuum conditions at least every eight hours. 
We point out that this is the only complete set of measurements of the FCS on ground in vacuum and hence the most representative (except for the flux).
In Figure~\ref{fig:dutv}, we show the integrated DU ready to be tested in the TV chamber.

\subsection{Data analysis tools}
\label{sec:swtools}
We acquired the SDD spectra with the DPPMCA display and acquisition software.
We then analyzed the spectra with custom Python scripts to fit the detected emission lines with a Gaussian+constant profile, which provides the estimate for $R_{SDD}$ (see Eq.~( \ref{eq:sddtogpdrate})). 
We acquired the CCD images with the Andor Solis software.
Finally, we acquired and analyzed the GPD measurements with the Python based ixpesw toolkit developed by the IXPE collaboration.

\section{Results}
\label{sec:results}
\subsection{SDD and CCD measurements results}
\label{sec:sddccdresults}
\begin{figure}[htbp]
\begin{center}
\begin{tabular}{c}
\includegraphics[height=7.5cm]{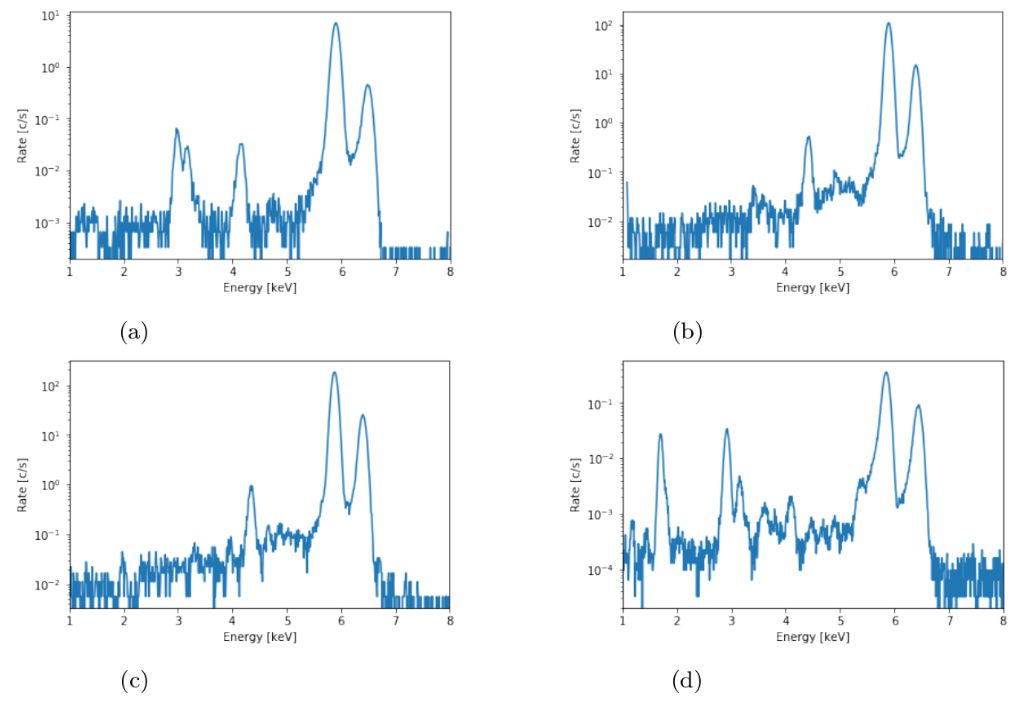}
\end{tabular}
\end{center}
\caption 
{ \label{fig:sddlog_FM02}
	Spectra of elements of FCS FM2 acquired with Amptek XR-100SDD SDD.
	(a) Spectrum of CalA2: the closely packed peaks of the Ag L$_{\alpha}$ and L$_{\beta}$ lines at 2.98 and 3.15 keV are visible, as are the 5.9 keV Mn K$_{\alpha}$ line and its escape at 4.2 keV and the K$_{\beta}$ line at 6.5 keV.
	(b) Spectrum of CalB2 and (c) spectrum of CalC2: the 5.9 keV Mn K$_{\alpha}$ line and its escape at 4.2 keV as well as the Mn K$_{\beta}$ line at 6.5 keV are visible in both spectra.
	(d) Spectrum of CalD: the lines at $\sim$3 keV due to argon in the air are well distinguished, similar to the 1.7 keV Si fluorescence.}
\end{figure}

\begin{figure}[htbp]
\begin{center}
\begin{tabular}{c}
\includegraphics[height=8cm]{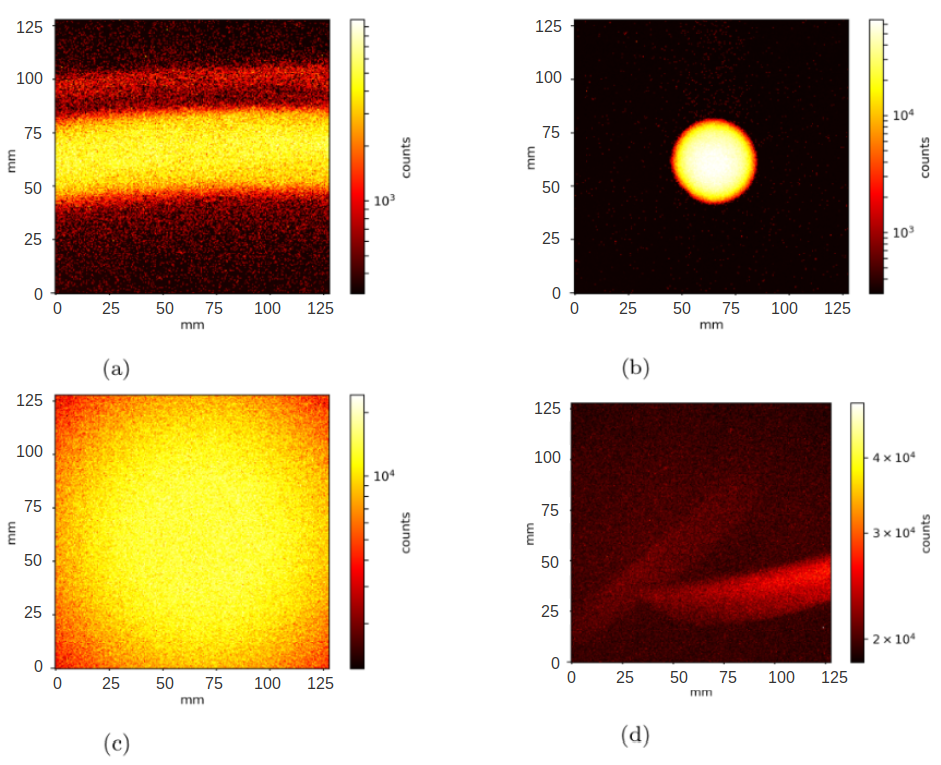}
\end{tabular}
\end{center}
\caption 
{ \label{fig:ccd_FM02}
Images of elements of FCS FM2 acquired with Andor iKon-M 934 CCD camera.
(a) CalA2: the central strip is due to the Bragg diffracted polarized photons at 3 and 5.9 keV, while the upper strip is interpreted as 6.5 keV photons diffracted at a different angle.
(b) CalB2: illuminates a 3 mm wide spot.
(c) CalC2: illuminates the whole detector.
(d) CalD2: due to heavy air absorption, the 1.7 keV fluorescence is not clear, instead two stripes are visible and interpreted as 5.9 and 6.5 keV photons diffracted by the Silicon target. } 
\end{figure}

In Figure~(\ref{fig:sddlog_FM02}), we show the spectra of the calibration sources of the FCS FM2, which is representative of all the models, as the others do not present significant differences. 
As the spectrum of CalA is acquired in air, the 2.98 and 3.15 keV silver fluorescence lines are polluted by the presence of the 2.96 and 3.19 keV lines due to the argon in the air. 
At an energy of 4.2 keV, the line arising due to the escape of the 5.9 keV K$_\alpha$ Mn line is evident in the spectra of CalA, CalB, and CalC . 
Due to the source geometry, this line is absent in the spectrum of CalD, whereas the argon lines are visible.
The flux of the 1.7 keV line in the spectrum of Cal D is heavily reduced by air absorption.
The spectra of CalB and CalC is dominated by the 5.9 keV Mn K$_{\alpha}$ and 6.5 keV Mn K$_{\beta}$ lines. 
In Table~(\ref{tab:results}), we report the expected GPD rate for the four FCSs obtained from Eq.~(\ref{eq:sddtogpdrate}) by considering the activity of the radioactive sources that will be employed on IXPE.
In Figure~(\ref{fig:ccd_FM02}), we show the images acquired by the CCD of the items of FCS FM2, as this set is sufficiently representative of the properties of the other three.
As expected by the Bragg diffraction theory, as seen in Fig.~(\ref{fig:diffrazionebragg}) and (\ref{fig:braggarccad}), the polarized diffracted photons at 3 and 5.9 keV appear as a curved strip across the detector surface.
Another strip above the main one is interpreted as 6.5 keV photons that are diffracted at a different angle. 
CalB illuminates a 3 mm wide spot, and CalC floods the entire detector surface.
CalD, must likewise illuminate the entire detector area; however, because the measurement is conducted in air, most of the 1.7 keV photons are absorbed. Moreover, only 5.9 and 6.5 keV bands, which are diffracted at the right angle by the silicon target, appear in the image as faint stripes. 
%
%
%
\subsection{Thermal vacuum results}
\label{sec:tv_results}
\begin{figure*}[htbp]
\begin{center}
\begin{tabular}{c}
\includegraphics[width=14.3cm]{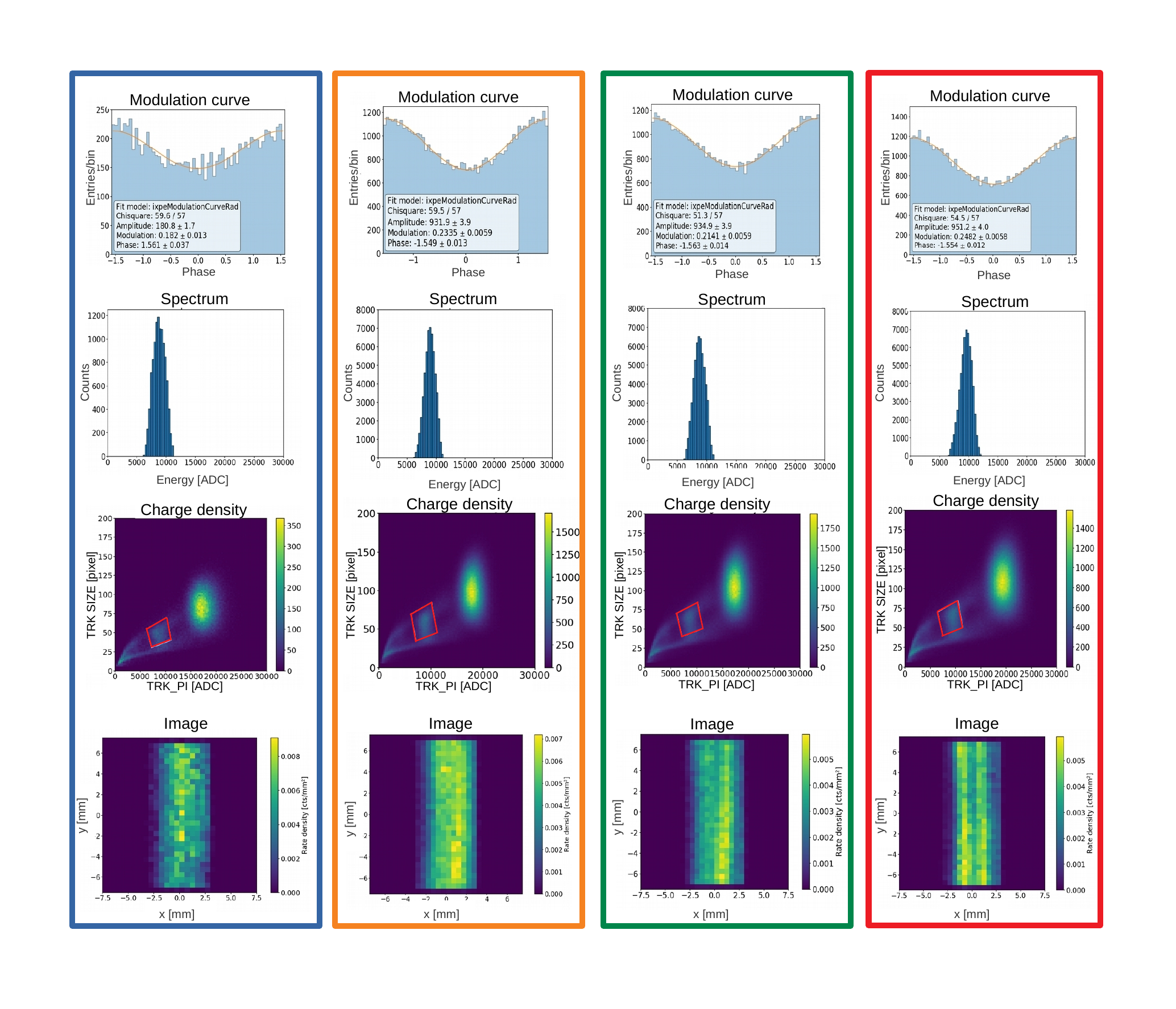} \\
(a) \hspace{3cm} (b) \hspace{3cm} (c) \hspace{3cm} (d)
\end{tabular}
\end{center}
\caption{Comparison of four CalA models for Ag fluorescence emission at 3 keV in TV. 
	From top to bottom: 
	modulation curve; 
	charge density-selected source spectrum; 
	charge density plot with the applied cut in energy and track size outlined in red; 
	image of the source on the GPD as rate density map for CalA1 (a), CalA2 (b), CalA3 (c), and CalA4 (d).}
\label{fig:CalA3}
\end{figure*}

\begin{figure*}[htbp]
	\begin{center}
		\begin{tabular}{c}
			\includegraphics[width=15cm]{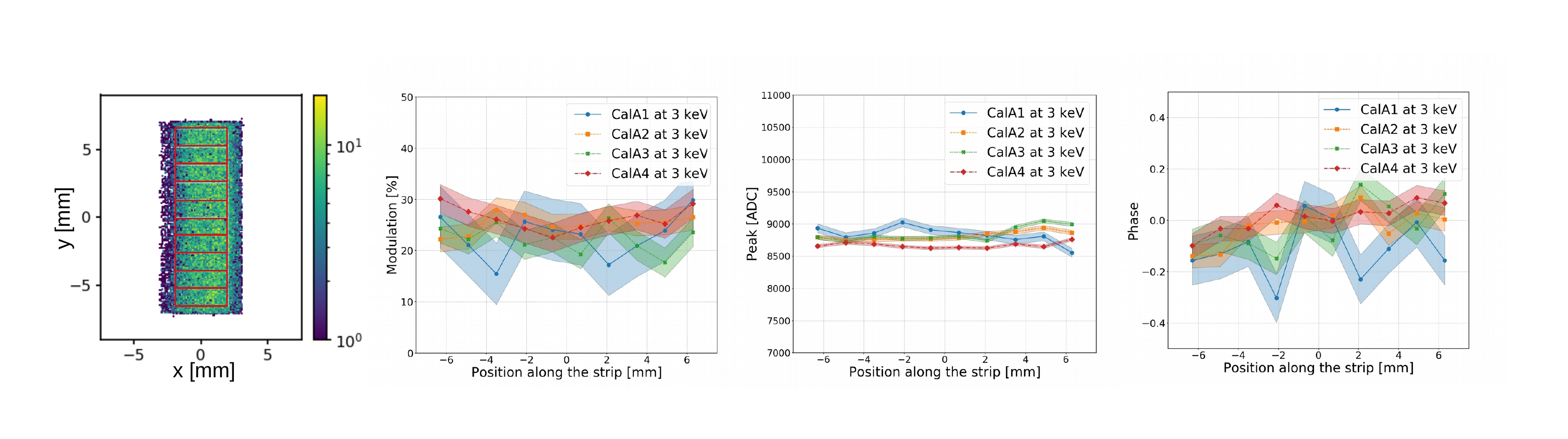} \\
			(a) \hspace{3cm} (b) \hspace{3cm} (c) \hspace{3cm} (d)
		\end{tabular}
	\end{center}
	\caption{Comparison of strip analysis of four CalA models for Ag fluorescence emission at 3 keV in TV.
		The shaded area represents the error in the measurement.
		 (a) Subdivisions of the strip for analysis highlighted in red; 
	(b) modulation as a function of position along the strip;
	(c) energy peak in ADC as a function of position along the strip;
	(d)	phase as a function of position along the strip.}
	\label{fig:CalA3strip}
\end{figure*}

\begin{figure*}[htbp]
	\begin{center}
		\begin{tabular}{c}
			\includegraphics[width=14.3cm]{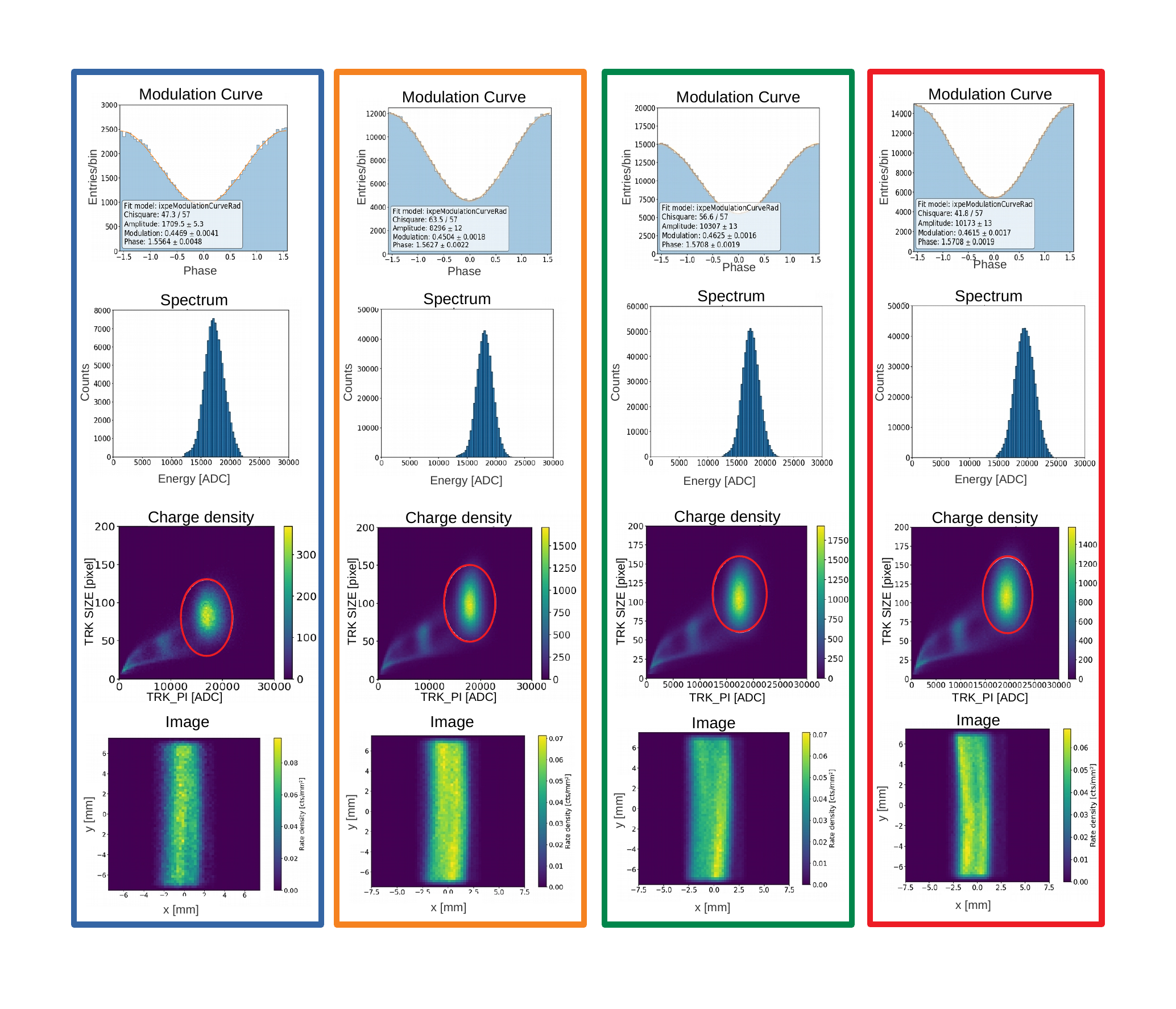} \\
			(a) \hspace{3cm} (b) \hspace{3cm} (c) \hspace{3cm} (d)
		\end{tabular}
	\end{center}
	\caption{Same as Fig.~(\ref{fig:CalA3}) but for the Mn emission at 5.9 keV in TV for CalA1 (a), CalA2 (b), CalA3 (c) and CalA4 (d).}
	\label{fig:CalA5}
\end{figure*}

\begin{figure*}[htbp]
	\begin{center}
		\begin{tabular}{c}
			\includegraphics[width=15cm]{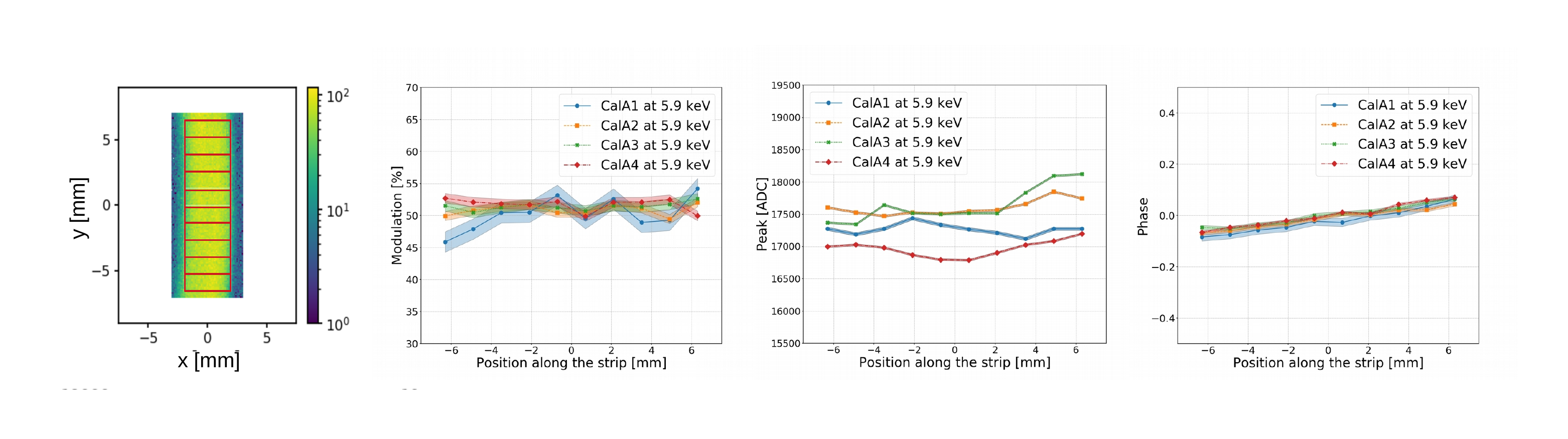} \\
			(a) \hspace{3cm} (b) \hspace{3cm} (c) \hspace{3cm} (d)
		\end{tabular}
	\end{center}
	\caption{Same as Fig.~(\ref{fig:CalA3}) but for the Mn emission at 5.9 keV in TV.}
	\label{fig:CalA5strip}
\end{figure*}

\begin{figure*}[htbp]
\begin{center}
\begin{tabular}{c}
\includegraphics[width=15cm]{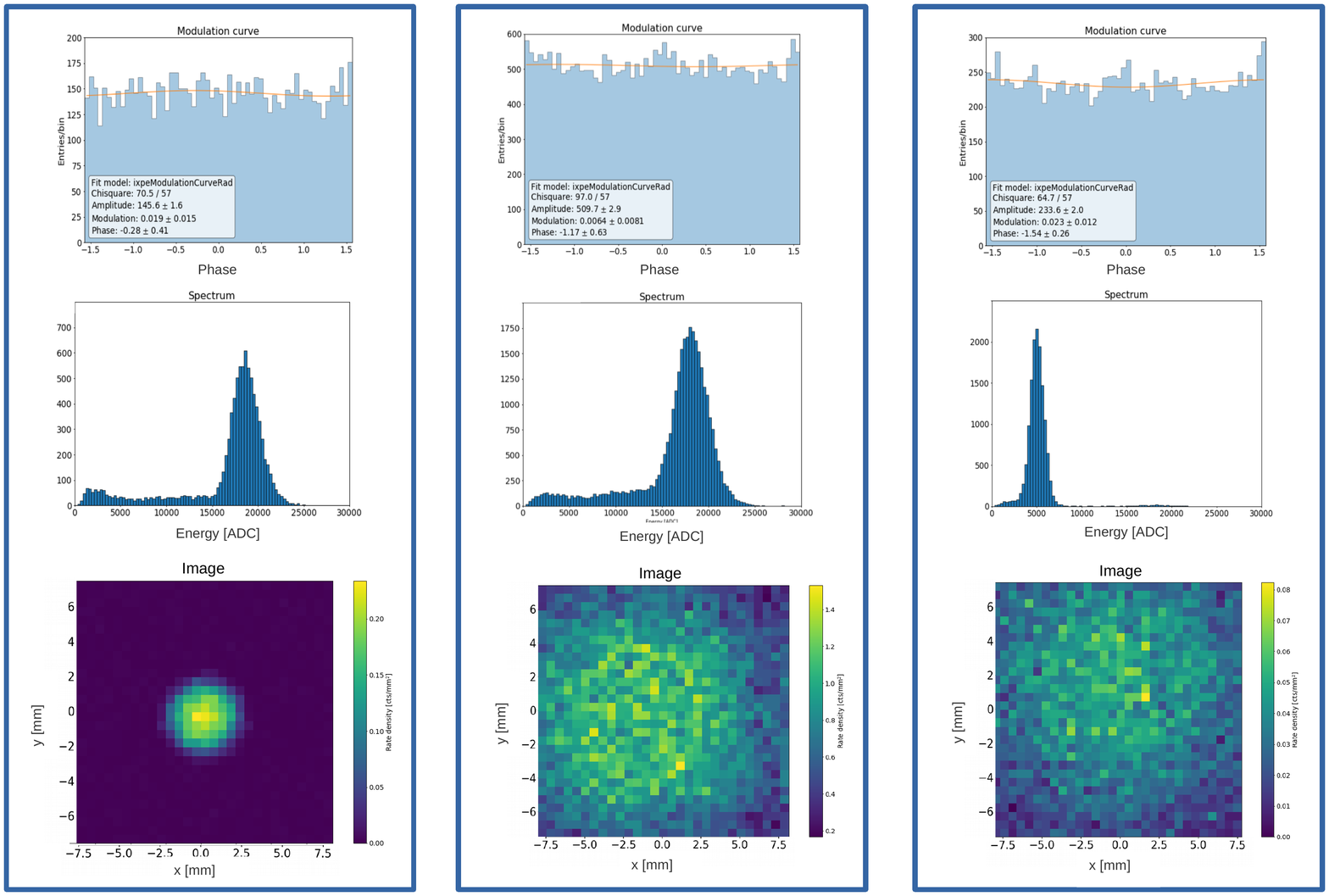} \\
(a) \hspace{4cm} (b) \hspace{4cm} (c)
\end{tabular}
\end{center}
\caption{Results for CalB2 (a), CalC2 (b), and CalD2 (c) in TV. 
		Top row: modulation curve;
		middle row: source spectrum;
		bottom row: image of source on GPD as rate density map.}
	\label{fig:CalBCD}
\end{figure*}

Because the polarized calibration source CalA is the most complex of the FCS, in Fig.~(\ref{fig:CalA3}) and (\ref{fig:CalA5}), we show the results of the four CalA in TV at 3 and 5.9 keV, respectively, side by side. 
In Figure~(\ref{fig:CalBCD}), we show the TV results of the monochromatic sources CalB, CalC, and CalD of the FCS FM2, which is representative, as there are no large differences across the four sets. 
For all the sources, we present in the first row the folded modulation curve, fitted with a $\cos^2$ function, showing to which extent the source is (un)polarized; in the second row the PHA spectrum; in the last row the source image on the detector as a rate density map, i.e., the image of the source on the detector in units of counts per second per mm$^2$, showing the uniformity of the illumination. \\
Unlike other sources, CalA is not a monochromatic source, and therefore its analysis is conducted separately for the 3 and 5.9 keV emission.
For CalA, we also show in the third row of Fig.~(\ref{fig:CalA3}) and (\ref{fig:CalA5}) the charge density map, i.e., the plane in which the photoelectron track size (defined as the number of triggered pixels of the photoelectron track above a detection threshold) and PHA of each event are represented. 
On this plane, we select a subset of events (shown as a red outline) associated to the emission line and remove the events generated close to the GPD beryllium window and to the gas electron multiplier, where they lose part of their energy in these passive materials. 
These events appear as thin strips above and below the spot of the events to be selected.
A $14 \times 6$ mm$^2$ region around the strip is selected to exclude from the analysis the events that occur on the edge of the detector and exhibit incomplete photoelectron tracks.
Finally, in Fig.~(\ref{fig:CalA3strip}) and (\ref{fig:CalA5strip}), we present the polarized calibration sources and the comparative analysis of the strip produced by the diffracted photons, at 3 and 5.9 keV, respectively: the strip is divided into ten rectangular boxes of 1.4 $\times$ 4 mm$^2$ regions. For each region the analog-to-digital converter (ADC) counts peak, the modulation and phase are plotted to verify their spatial uniformity.
These CalA results are corrected for the spatial differences in gain of the detector and by the GPD response to unpolarized radiation, both of which are characteristics unique to each detector and that have been determined during the characterization campaign of the DUs. \\
In Table~(\ref{tab:results}), we summarize the results of the measurements in TV for all four FCS: in particular, we list the flight rate with and without dead time correction (within parentheses), the MDA reached, the measured modulation amplitude, and the polarization angle phase.
%
%
%
%
%
%
%
%
\section{Discussion}
\label{sec:discussion}
Table~(\ref{tab:results}) lists and compares the results of the test with SDD and with the DU in TV.
The results obtained using the DU in TV are presented with and without (within parentheses) the dead time correction.
Because the dead time of the SDD is almost negligible, the results of the extrapolation from the SDD must be compared with the ones from GPD corrected for the dead time.
The dead time corrected rate is important to establish the real activity of the radioactive sources.
The correction is particularly important for CalC, which is a bright source.
The raw counting rates without dead time correction are used when dealing with the statistical significance of a measurement.
The extrapolated GPD rate from the SDD measurements is in good agreement with the rates measured in TV with the GPD in the case of CalA at 5.9 keV, as well as in the cases of CalB, and CalD.
The rates of CalC are overestimated because the geometric correction assumes a perfectly uniform illumination, while the images with CCD and GPD show that the central region is more illuminated than the corners of the detector. 
Moreover, because the real activity of the source used during the measurement is unknown, it was possibly above the nominal value.
The rates of CalA at 3 keV appear to be higher in the SDD spectra because of the argon contamination.
The systematic underestimation in the expected and measured rates of CalD is explained by the difficulty in correctly estimating the effect of air absorption at 1.7 keV, which removes most of the photons. 
We are therefore confident that we will obtain the required statistical significance from the flight activities. \\
The flight counting rates extrapolated by the TV measurements are consistently larger than the requirements listed in Table~(\ref{tab:requirements}), with the differences across the sets that can be ascribed to the different energy resolution of each detector.
The images of the source on the GPD (e.g., Fig.~(\ref{fig:CalA3}),~(\ref{fig:CalA5}) and (\ref{fig:CalBCD}) for the FCS FM2) are compatible with the one observed with the CCD (Fig.~(\ref{fig:ccd_FM02})).
The diffraction stripes observed in CalD with the CCD can easily be removed in the analysis phase by applying a cut in energy, such that their impact on the calibrations is null. \\
In Figure~(\ref{fig:CalA3strip}) and (\ref{fig:CalA5strip}), we show and compare across the four FCS the analysis of the Bragg arc generated by the polarized calibration sources: the energy response of the detector, traced by the value of the peak in ADC, and the modulation amplitude are constant across the strip, with fluctuation small enough to be negligible during the in-flight calibrations.
Moreover, the value of the modulation, the peak position in energy, and the phase are in good agreement across the four sets, especially in the central region of the GPD.
As expected from the theory of Bragg diffraction, the polarization angle, traced by the phase, stays tangent to the Bragg arc, with the observed change in the sign due to the choice of the reference frame, with zero at the center of the detector, where the polarization changes direction.
The four CalA sources show a modulation amplitude that is consistent with radiation polarized at 67\% and 69\% at 3 and 5.9 keV, respectively, as expected from \cite{1993Henke}. \\
The unpolarized collimated sources, CalB, show modulations that are consistent with no polarization.
The unpolarized calibration sources CalB, CalC, and CalD show modulations consistent with no polarization, except for the case of CalD3, where a 6\% modulation amplitude is observed with a 3$\sigma$ significance level. \\
However, CalD will be mostly employed to monitor the gain, and only partially to check for the presence of spurious modulation. \\
IXPE will orbit at 600 km of altitude, with an orbital period of 5.76 ks and 35\% of this time affected by Earth occultations of X-ray targets.
This fraction of the orbital period can be used for calibration purposes: during $\sim$2 ks long sessions, the three DUs will be calibrated one per orbit to guarantee that at least two DUs will remain operative and reduce the risk associated with a failure of the FCW mechanism.
Moreover, this strategy allows to limit the count rate for the effective downlink.
Eventually, all three DUs will be exposed to all four calibration sources.
These calibrations will monitor potential changes in the detector gain, modulation factor, polarization response, and, within the limits of statistics, spurious modulation. 
The results will be cross-checked with the one performed on the ground, further advancing our understanding of the technology of the GPD and investigating possible secular variations of physical parameters. \\
In Table~(\ref{tab:results}), we also report the number of orbits necessary to reach an MDA of 2\%, corresponding to an absolute error of 1\% in the determination of the modulation amplitude, as in \cite{2013Strohmayer} the uncertainty on the measurement of the amplitude is given by MDA/2. \\
This means that the calibration of the DU with CalA at 3 keV on four orbits, lasting a total of a three hours, would allow us to reach a level of 22 sigmas at 3 keV and 45 sigmas at 5.9 keV. 
Finally, the measured modulation is convoluted with the GPD modulation factor to obtain the polarimetric sensitivity. 
To determine these values, the mean raw rate from TV measurements without cuts and dead time correction is used. 

\begin{table}[htbp]\scriptsize
\caption{Comparison of results of SDD tests with the ones performed with GPD in TV on the four FM of the FCS.
		The results of CalA at 3 keV marked with an * are affected by argon contamination.
		The results within parentheses denote the rates with dead time correction. 
		The last two columns show the mean time needed to acquire enough counts, based on the mean rate from each source, to reach an absolute error of 1\% on the modulation amplitude and the number of IXPE orbits needed (assuming that calibrations are performed within the $35\%$ of the orbit during Earth occultation of celestial sources).}
\label{tab:results}     
\begin{center}
\begin{tabular}{|l|l|l|l|l|l|l|l|l|}
	\hline
	\rule[-1ex]{0pt}{3.5ex}
	Source					& Energy			& Flight rate 					& Flight rate 	& MDA	& Modulation	& Phase	&Mean time to	& Orbits to	\\
	\rule[-1ex]{0pt}{3.5ex}
	& [keV]				& from SDD						& from DU in TV	& [\%]	& [\%]			&		& 1\% absolute error	& 1\% absolute error\\
	\rule[-1ex]{0pt}{3.5ex}
	&					& [c/s]							& [c/s]			& 		&				&		&[ks]		& 			\\
	\hline\hline
	\rule[-1ex]{0pt}{3.5ex}
	\multirow{2}{*}{CalA1}	&\multirow{8}{*}{3}	& \multirow{2}{*}{5.5*$\pm$0.1}& 3.03$\pm$0.04	&\multirow{2}{*}{2.8}  	&\multirow{2}{*}{18.2$\pm$1.3}	&\multirow{2}{*}{1.56$\pm$0.04}	&\multirow{8}{*}{11.488}	&\multirow{8}{*}{4.34}\\	
	\rule[-1ex]{0pt}{3.5ex}
	&  					&								&(3.05$\pm$0.04)&		&				&		&			&\\		
	\cline{1-1}\cline{3-7}
	\rule[-1ex]{0pt}{3.5ex}
	\multirow{2}{*}{CalA2}	& 					& \multirow{2}{*}{5.5$\pm$0.1} 	& 3.56$\pm$0.02 &\multirow{2}{*}{2.2}&\multirow{2}{*}{23.4$\pm$0.6}	&\multirow{2}{*}{-1.55$\pm$0.01}	&	&\\
	\rule[-1ex]{0pt}{3.5ex}
	& 				 	&								&(3.58$\pm$0.02)&		&				&		&	&\\
	\cline{1-1}\cline{3-7}
	\rule[-1ex]{0pt}{3.5ex}
	\multirow{2}{*}{CalA3}	& 					& \multirow{2}{*}{5.7$\pm$0.1} 	& 3.34$\pm$0.02	&\multirow{2}{*}{2.2}  & \multirow{2}{*}{21.4$\pm$0.6}& \multirow{2}{*}{-1.56$\pm$0.01}	&	&\\
	\rule[-1ex]{0pt}{3.5ex}
	&  					&								&(3.40$\pm$0.02)&		&			&		&	&\\		
	\cline{1-1}\cline{3-7}
	\rule[-1ex]{0pt}{3.5ex}
	\multirow{2}{*}{CalA4}	& 					& \multirow{2}{*}{5.4$\pm$0.1}	& 3.42$\pm$0.02	&\multirow{2}{*}{2.1}  &\multirow{2}{*}{24.8$\pm$0.6}&\multirow{2}{*}{-1.55$\pm$0.01}	&	&\\
	\rule[-1ex]{0pt}{3.5ex}
	&  					&								&(3.47$\pm$0.02)&		&			&		&	&\\		
	\hline
	\rule[-1ex]{0pt}{3.5ex}
	\multirow{2}{*}{CalA1}	& \multirow{8}{*}{5.9}& \multirow{2}{*}{40.2$\pm$0.2}& 40.3$\pm$0.1	&\multirow{2}{*}{0.9}  	&\multirow{2}{*}{44.7$\pm$0.4}&\multirow{2}{*}{1.536$\pm$0.005}	&\multirow{8}{*}{0.932}	& \multirow{8}{*}{0.35}\\
	\rule[-1ex]{0pt}{3.5ex}
	&  					&								&(40.6$\pm$0.1)	&		&		&	&			&\\	
	\cline{1-1}\cline{3-7}
	\rule[-1ex]{0pt}{3.5ex}
	\multirow{2}{*}{CalA2}	&					& \multirow{2}{*}{45.6$\pm$0.4}	& 41.0$\pm$0.1	& \multirow{2}{*}{0.7}  	& \multirow{2}{*}{45.0$\pm$0.2}&\multirow{2}{*}{1.563$\pm$0.002}	&			&\\
	\rule[-1ex]{0pt}{3.5ex}
	&  					&								&(41.5$\pm$0.1)	&		&		&	&			&\\	
	\cline{1-1}\cline{3-7}
	\rule[-1ex]{0pt}{3.5ex}
	\multirow{2}{*}{CalA3}	&					& \multirow{2}{*}{48.7$\pm$0.4}	& 42.0.6$\pm$0.1	& \multirow{2}{*}{0.6}  	& \multirow{2}{*}{46.3$\pm$0.2}&\multirow{2}{*}{1.571$\pm$0.002}	&			&\\
	\rule[-1ex]{0pt}{3.5ex}
	&  					&								&(42.5$\pm$0.1)	&		&		&	&			&\\	
	\cline{1-1}\cline{3-7}
	\rule[-1ex]{0pt}{3.5ex}
	\multirow{2}{*}{CalA4}	&					& \multirow{2}{*}{44.7$\pm$0.4}	& 41.2$\pm$0.1	&\multirow{2}{*}{0.6}  	&\multirow{2}{*}{46.2$\pm$0.2}&\multirow{2}{*}{1.570$\pm$0.002}	&			&\\
	\rule[-1ex]{0pt}{3.5ex}
	&  					&								&(42.1$\pm$0.1)	&		&		&	&			&\\	
	\hline 
	\rule[-1ex]{0pt}{3.5ex}
	\multirow{2}{*}{CalB1}	& \multirow{8}{*}{5.9}& \multirow{2}{*}{70$\pm$1}	& 65$\pm$2		&\multirow{2}{*}{13.4}	&\multirow{2}{*}{7.7$\pm$4.3}&\multirow{2}{*}{0.5$\pm$0.3}	&\multirow{8}{*}{0.642}	& \multirow{8}{*}{0.24} \\
	\rule[-1ex]{0pt}{3.5ex}
	&  					&								&(71$\pm$2)	&		&		&	&			&\\	
	\cline{1-1}\cline{3-7}
	\rule[-1ex]{0pt}{3.5ex} 
	\multirow{2}{*}{CalB2}	&					& \multirow{2}{*}{79.7$\pm$0.9}	& 71.4$\pm$0.8		&\multirow{2}{*}{4.7}	&\multirow{2}{*}{1.9$\pm$1.5}&\multirow{2}{*}{-0.3$\pm$0.4}	&&\\
	\rule[-1ex]{0pt}{3.5ex} 
	&  					&								&(76.3$\pm$0.8)		&	& &	& &\\	
	\cline{1-1}\cline{3-7}
	\rule[-1ex]{0pt}{3.5ex} 
	\multirow{2}{*}{CalB3}&						& \multirow{2}{*}{83$\pm$1}	& 72.8$\pm$0.8		&\multirow{2}{*}{5.0}	&\multirow{2}{*}{0.6$\pm$1.6}&\multirow{2}{*}{-1.2$\pm$1.4}	& &\\
	\rule[-1ex]{0pt}{3.5ex} 
	&  					&								&(78.5$\pm$0.9)	&	& &	& &\\	
	\cline{1-1}\cline{3-7}
	\rule[-1ex]{0pt}{3.5ex} 
	\multirow{2}{*}{CalB4}	&					& \multirow{2}{*}{83$\pm$1}	& 77.8$\pm$0.9		&\multirow{2}{*}{5.0}	&\multirow{2}{*}{0.5$\pm$1.6}&\multirow{2}{*}{-0.7$\pm$1.5} 	&&\\
	\rule[-1ex]{0pt}{3.5ex} 
	&					&  								&(83$\pm$1)		&	& &	&&\\	
	\hline
	\rule[-1ex]{0pt}{3.5ex} 
	\multirow{2}{*}{CalC1}	& \multirow{8}{*}{5.9}& \multirow{2}{*}{256.6$\pm$0.7}& 131.7$\pm$0.6&\multirow{2}{*}{1.9}	&\multirow{2}{*}{0.8$\pm$0.6}&\multirow{2}{*}{0.8$\pm$0.4} &\multirow{8}{*}{0.300} & \multirow{8}{*}{0.11} \\
	\rule[-1ex]{0pt}{3.5ex} 
	& 				 	&								&(170.2$\pm$0.7)	&	& & &	&\\	
	\cline{1-1}\cline{3-7}
	\rule[-1ex]{0pt}{3.5ex} 
	\multirow{2}{*}{CalC2}	& 					& \multirow{2}{*}{279$\pm$1}	& 150.0$\pm$0.4 &\multirow{2}{*}{1.2}	&\multirow{2}{*}{0.6$\pm$0.8}&\multirow{2}{*}{-1.2$\pm$0.6} & &\\
	\rule[-1ex]{0pt}{3.5ex} 
	& 				 	&								&(192.8$\pm$0.5)&	& &	&						&\\	
	\cline{1-1}\cline{3-7}
	\rule[-1ex]{0pt}{3.5ex} 
	\multirow{2}{*}{CalC3}	& 					& \multirow{2}{*}{271$\pm$1}	& 177.7$\pm$0.7	&\multirow{2}{*}{1.6}	&\multirow{2}{*}{1.8$\pm$0.5}&\multirow{2}{*}{-1.5$\pm$0.1}	 & &\\
	\rule[-1ex]{0pt}{3.5ex} 
	&  					&								&(225.7$\pm$0.8)&	& &	&						&\\	
	\cline{1-1}\cline{3-7}
	\rule[-1ex]{0pt}{3.5ex} 
	\multirow{2}{*}{CalC4}	& 					& \multirow{2}{*}{276$\pm$1}	&154.9$\pm$0.5	&\multirow{2}{*}{1.4}	&\multirow{2}{*}{0.3$\pm$0.5}	&\multirow{2}{*}{0.8$\pm$0.8} & &\\
	\rule[-1ex]{0pt}{3.5ex} 
	&  					&								&(192.7$\pm$0.6)&	& &	&						&\\	
	\hline
	\rule[-1ex]{0pt}{3.5ex}
	\multirow{2}{*}{CalD1}	& \multirow{8}{*}{1.7}& \multirow{2}{*}{89.4$\pm$0.2}& 105.8$\pm$0.7	& \multirow{2}{*}{2.7} & \multirow{2}{*}{1.1$\pm$0.9} & \multirow{2}{*}{-0.5$\pm$0.4} & \multirow{8}{*}{0.408}& \multirow{8}{*}{0.15}\\
	\rule[-1ex]{0pt}{3.5ex} 
	&  					&								&(106.8$\pm$0.7)&	&	&	&			&\\	
	\cline{1-1}\cline{3-7}
	\rule[-1ex]{0pt}{3.5ex}
	\multirow{2}{*}{CalD2}	&					& \multirow{2}{*}{84.6$\pm$0.2}& 113$\pm$1		& \multirow{2}{*}{3.6}	& \multirow{2}{*}{2.3$\pm$1.2}	& \multirow{2}{*}{-1.5$\pm$0.3} & &\\
	\rule[-1ex]{0pt}{3.5ex} 
	&  					&								&(114$\pm$1)	&	& &	&						&\\	
	\cline{1-1}\cline{3-7}
	\rule[-1ex]{0pt}{3.5ex} 
	\multirow{2}{*}{CalD3}	&					& \multirow{2}{*}{101.0$\pm$0.4}& 121$\pm$1		& {3.6}	& {6.0$\pm$2.0}	& \multirow{2}{*}{-1.3$\pm$0.1} & &\\
	\rule[-1ex]{0pt}{3.5ex} 
	&  					&								&(122$\pm$1)	&	& &	&						&\\	
	\cline{1-1}\cline{3-7}
	\rule[-1ex]{0pt}{3.5ex}
	\multirow{2}{*}{CalD4}	&					& \multirow{2}{*}{103.5$\pm$0.2}& 111$\pm$2		& \multirow{2}{*}{6.7}	& \multirow{2}{*}{3.2$\pm$2.3}	& \multirow{2}{*}{-1.4$\pm$0.4} & &\\
	\rule[-1ex]{0pt}{3.5ex} 
	&  					&								&(112$\pm$2)	&	& &	&						&\\	
	
	\hline
\end{tabular}
\end{center}
\end{table}
%
\section{Conclusion}
\label{sec:conclusion}
The items of the FCS were tested first with a commercial SDD and CCD to verify their operation. 
The calibration sources were then tested in TV with the flight DU to derive their spectra, images on the detectors, and polarimetric performance.
The morphology of the sources, studied independently with CCD and GPD, are consistent. 
The expected counting rates are comparable across the different FMs, with differences that can be ascribed to the different energy resolution of each DU.
The counting rates satisfy the requirements, and the modulation of the polarized sources is consistent with the one expected from Bragg diffraction. \\ 
Three FCS will be installed on the three IXPE DUs, and one will be installed in a fourth DU that acts as a spare and serves for off-line testing before and after the launch.
With the expected launch date of April 2021, due to the decay of the radioactive sources, the reported rates will be reduced to 77.6\% of their initial values, without significantly impacting the in-orbit calibration times.
However, owing to the ongoing Covid-19 pandemic, the launch may be delayed until September 2021.
In this case, by that time, the rate of the flight radioactive sources will have decayed to 71.1\% of their initial value.
IXPE is expected to last at least two years, such that by the end of the mission lifetime, the activity of the sources will be reduced to $\sim$ 40\%.
During the mission, the FCS will help validate the scientific results of IXPE by checking the detector response to point-like and extended sources. 
In summary, the results obtained on-ground, when extrapolated to the ones expected in flight, provide us with confidence that the FCS will be able to properly monitor the performance of the DUs.

\subsection*{Disclosures}
The authors have no relevant financial interests with regard to the publication of this manuscript and no other conflicts of interest to disclose.

\acknowledgments 
The Italian contribution to the IXPE mission is supported by the Italian Space Agency through agreement ASI-INAF n.2017-12-H.0. \\
INAF and INFN are significantly contributing through internal resources of manpower, structures and equipment. \\
The FCW was manufactured by OHB Italia SpA and includes the heritage from eROSITA experiment aboard the Spectrum X Gamma Mission. \\
Radioactive sources were procured from Eckert \& Ziegler. \\
We thank the anonymous reviewers for their appreciations and useful comments.

\bibliographystyle{spiejour}   
\bibliography{biblio.bib}   


\vspace{2ex}\noindent\textbf{Riccardo Ferrazzoli} is a PhD student of the XXXIV cycle conducting his PhD in Astronomy, Astrophysics and Space Science at INAF-IAPS.
He received his BS and MS degrees in Physics and Astrophysics from the University of Roma Tor Vergata in 2013 and 2016, and a Master in Space Science and Technology in 2018.
He is a science participant in the IXPE mission.
%


\end{document}